\documentclass[10pt]{iopart}
\usepackage{iopams}
\usepackage{cite}

\begin{document}

\title[Is a phonon ... a Goldstone boson?]{Is a phonon excitation of a superfluid Bose gas \\ a Goldstone boson?}

\author{Maksim Tomchenko}

\address{Bogolyubov Institute for Theoretical Physics, 14b,  Metrolohichna Str., Kyiv 03143, Ukraine}
\ead{mtomchenko@bitp.kyiv.ua} \vspace{10pt}
\begin{indented}
\item[]January 2026
\end{indented}

\begin{abstract}
It is generally accepted that phonons in a superfluid Bose gas are
Goldstone bosons. This is justified by spontaneous symmetry breaking
(SSB), which is usually defined as follows: the Hamiltonian of the
system is invariant under the
$U(1)$ transformation $\hat{\Psi}(\mathbf{r},t)\rightarrow e^{i\alpha}%
\hat{\Psi}(\mathbf{r},t)$, whereas the order parameter
$\Psi(\mathbf{r},t)$ is not. However, the strict definition of SSB
is different: the Hamiltonian and the boundary conditions are
invariant under a symmetry transformation, while the \emph{ground
state} is not. Based on the latter criterion, we study a finite
system of spinless, weakly interacting bosons using three
approaches: the standard Bogoliubov method, the
particle-number-conserving Bogoliubov method, and the approach based
on the exact ground-state wave function. Our results show that the
answer to the question in the title is \textquotedblleft
no\textquotedblright. Thus, phonons in a real-world (finite)
superfluid Bose gas are similar to sound in a classical gas: they
are not Goldstone bosons, but quantised collective vibrational modes
arising from the interaction between atoms. In the case of an
infinite Bose gas, however, the picture becomes paradoxical: the
ground state can be regarded as either infinitely degenerate or
non-degenerate, making the phonon both similar to a Goldstone boson
and different from it.
\end{abstract}

%
\vspace{2pc} \noindent{\it Keywords}: Bose gas, spontaneous symmetry
breaking, phonon, Goldstone boson

%
%
\ioptwocol

\section{Introduction}
It is widely believed that sound waves (phonons) in superfluid
($T<T_{\lambda}$) Bose gases and liquids are Goldstone bosons
\cite{anderson1984,miransky1994,forster2018,wezel2019,powell2020}
(recall that $T_{\lambda}$ is the transition temperature to the
superfluid state). On the other hand, phonons in the same gas
(liquid) at $T>T_{\lambda}$ are no longer Goldstone bosons, but
classical sound waves that exist due to the interaction of atoms
with each other. This picture seems somewhat strange, since the
interaction between atoms at $T<T_{\lambda}$ is exactly the same as
at $T>T_{\lambda}$. Moreover, the single-phonon wave function of the
Bose gas is invariant under permutations of atoms, so that a phonon
is equally created by the motion of \textit{all} atoms, both those
that are in the condensate and those that are
not~\cite{fc1956,fey1972,yuv2}. In what follows we will try to find
out whether phonons really become Goldstone bosons at
$T<T_{\lambda}$. This is important for understanding the nature of
the sound mode in superfluid systems and of the superfluidity
phenomenon itself.

A similarity between the nature of phonons  and that of Goldstone
bosons was substantiated in two ways. The simplest is to show that
the second quantised Hamiltonian (Lagrangian, free energy) of a
many-particle system is invariant
under the $U(1)$-rotation $\hat{\Psi}(\mathbf{r},t)\rightarrow e^{i\alpha}%
\hat{\Psi}(\mathbf{r},t)$ (for the order parameter, $\Psi(\mathbf{r}%
,t)\rightarrow e^{i\alpha}\Psi(\mathbf{r},t)$), while the order
parameter $\Psi(\mathbf{r},t)$ is not invariant.  This property was
interpreted as a spontaneous breakdown of the continuous $U(1)$
symmetry, from whence authors concluded, by appealing to the theorem
of J.~Goldstone \cite{goldstone1961,goldstone1962,coleman1973}, that
there is a massless boson in the system. For a quantum-mechanical
system, such a boson corresponds to a
phonon~\cite{miransky1994,powell2020}. In this approach, the free
energy (or Lagrangian) is written phenomenologically and it is
postulated that the order parameter and the condensate are the same.

More rigorous approaches introduce the order parameter as the
average
$\langle0|\hat{\Psi}(\mathbf{r},t)|0\rangle\equiv\Psi(\mathbf{r},t)$
over the ground state $|0\rangle$ of the system
\cite{anderson1984,miransky1994,forster2018,wezel2019}. However,
according to the
second-quantisation formalism, $\langle0|\hat{\Psi}(\mathbf{r}%
,t)|0\rangle=0$. Therefore, either a non-zero order parameter
$\langle 0|\hat{\Psi}(\mathbf{r},t)|0\rangle$ is
postulated~\cite{anderson1966}, or the quasi-average
$\langle0|\hat{\Psi }(\mathbf{r},t)|0\rangle_{q}$ is
used~\cite{bogquasi1,bogquasi2} instead of the standard average, or
the state $|0\rangle$ is considered to be a state with an indefinite
number of
particles~\cite{miransky1994,forster2018,wezel2019,hohenberg1965}.

There are two definitions of spontaneous symmetry breaking (SSB):
statistical and quantum. The statistical definition is as follows:
SSB is present if the Hamiltonian of a system is invariant under
some symmetry, but the statistical equilibrium state is
not~\cite{bogquasi1,bogquasi2}. This definition is usually
applicable only to infinite
systems~\cite{wezel2019,bogquasi1,bogquasi2}. The quantum definition
is applicable to both finite and infinite systems. Since we are
interested in real-world systems, which are always finite, we will
use the quantum definition: SSB is realised when the Lagrangian (or
Hamiltonian) of an infinite system is invariant under a symmetry
transformation, but the ground state is
not~\cite{miransky1994,coleman}. For a finite system, the symmetry
of boundary conditions (BCs) can be lower than that of the
Hamiltonian (Lagrangian). It is clear that in this case the symmetry
of the ground state is determined by the BCs and is lower than the
symmetry of the Hamiltonian; however, this is not SSB. Therefore,
the strict definition of SSB is as follows~\cite{sgs}: the boundary
value problem (the Hamiltonian and the BCs) is invariant under some
symmetry, but the ground state is not. The quantum definition is
formally a particular case ($T=0$) of the statistical one;
however, we treat them separately because their application requires
different mathematical frameworks.

According to the Goldstone theorem, in the case of spontaneous
breaking of \textit{continuous} symmetry, there must be a massless
boson in the system
\cite{goldstone1961,goldstone1962,miransky1994,coleman}. This
theorem was proved in the quantum field theory, where the ground
state is a state without \textit{particles}, although particles can
in principle be created and destroyed. However, we consider a
quantum-mechanical system. In this case, the ground state is a state
without \textit{quasiparticles}, and particles cannot be created or
destroyed. The Goldstone theorem is inapplicable to
quantum-mechanical systems. Therefore, strictly speaking, the answer
to the question in the title of this article is always negative.

A quantum-mechanical analog of the Goldstone theorem might sound
like this: if SSB is exhibited in a many-particle system, then there
must be elementary quasiparticles with a gapless dispersion law.
Such a theorem has been proved for an infinite system (the
$1/q^{2}$-theorem \cite{bogquasi1,bogquasi2}) but not for a finite
one. However, we can assume that it is also true for a finite system
(note that SSB is possible for a finite system, in particular for a
system of spins or multipoles). In the case of a system of spinless
bosons, a phonon can be considered as \emph{an analogue} of the
Goldstone boson, if the Hamiltonian and BCs are
invariant under the $U(1)$ transformation $\hat{\Psi}(\mathbf{r}%
,t)\rightarrow e^{i\alpha}\hat{\Psi}(\mathbf{r},t)$, but the
ground-state wave function is not.

In this paper, we investigate this problem in such a rigorous
approach, using three different models of a Bose system (section 2).
This analysis allows us to give an explicit answer to the question
of whether the phonon in a finite superfluid Bose system is an
analogue of the Goldstone boson. In section 3, we explore the origin
of the SSB in an infinite Bose system. Section 4 contains a final
discussion.

\section{Does spontaneous breaking of $U(1)$ symmetry occur in a finite
system of interacting spinless bosons?}

\subsection{The standard Bogoliubov approach}

Let us study the problem in three different ways. We start with the
most famous one, the Bogoliubov model \cite{bog1947}. The SSB
problem has already been studied in this approach in the
book~\cite{miransky1994}. We
will reproduce this analysis for the general potential $U(|\mathbf{r}%
_{j}-\mathbf{r}_{l}|)$, while in~\cite{miransky1994} the point-like
potential $U(|\mathbf{r}_{j}-\mathbf{r}_{l}|)=U_{0}\delta(\mathbf{r}%
_{j}-\mathbf{r}_{l})$ was considered.

Consider a three-dimensional (3D) system of weakly interacting
spinless bosons with periodic BCs. The exact second-quantised Hamiltonian%
\begin{eqnarray}
&&\hat{H}=
-\frac{\hbar^2}{2m}\int\limits_{V}d\mathbf{r}\hat{\psi}^{+}(\mathbf{r},t)\triangle
\hat{\psi}(\mathbf{r},t) \label{1}  \\ &&+
\frac{1}{2}\int\limits_{V}d\mathbf{r}
d\mathbf{r}^{\prime}U(|\mathbf{r}-\mathbf{r}^{\prime}|)
\hat{\psi}^{+}(\mathbf{r},t)\hat{\psi}^{+}(\mathbf{r}^{\prime},t)
\hat{\psi}(\mathbf{r},t)\hat{\psi}(\mathbf{r}^{\prime},t) \nonumber
\end{eqnarray}
is reduced to the approximate Bogoliubov Hamiltonian%
\begin{eqnarray}
&&\hat{H}_{\mathrm{app}}=\frac{N^{2}\nu(0)}{2V}+\sum\limits_{\mathbf{k}\neq
0}\frac{\hbar^{2}k^{2}}{2m}\hat{a}_{\mathbf{k}}^{+}\hat{a}_{\mathbf{k}}%
\nonumber \\ &&+\sum\limits_{\mathbf{k}\neq0}\frac{\nu(\mathbf{k})}{2V}\left[  a_{0}^{2}%
\hat{a}_{\mathbf{k}}^{+}\hat{a}_{-\mathbf{k}}^{+}+(a_{0}^{\ast})^{2}\hat
{a}_{\mathbf{k}}\hat{a}_{-\mathbf{k}}+2a_{0}^{\ast}a_{0}\hat{a}_{\mathbf{k}%
}^{+}\hat{a}_{\mathbf{k}}\right]  \label{2}%
\end{eqnarray}
when the following formulae are used: $\hat{a}_{0}\approx a_{0}\gg1$,%
\begin{equation}
\hat{\psi}(\mathbf{r},t)=\frac{1}{V^{1/2}}\sum\limits_{\mathbf{k}}\hat
{a}_{\mathbf{k}}e^{i\mathbf{k}\mathbf{r}}, \label{3}%
\end{equation}%
\begin{equation}
\hat{\psi}^{+}(\mathbf{r}%
,t)=\frac{1}{V^{1/2}}\sum\limits_{\mathbf{k}}\hat{a}_{\mathbf{k}}%
^{+}e^{-i\mathbf{k}\mathbf{r}}, \label{3p}%
\end{equation}%
\begin{equation}
\hat{N}=\sum\limits_{\mathbf{k}}\hat{a}_{\mathbf{k}}^{+}\hat{a}_{\mathbf{k}%
}\approx N\approx a_{0}^{\ast}a_{0}+\sum\limits_{\mathbf{k}\neq0}\hat
{a}_{\mathbf{k}}^{+}\hat{a}_{\mathbf{k}}\approx a_{0}^{\ast}a_{0}=N_{0},
\label{4}%
\end{equation}%
\begin{equation}
U(|\mathbf{r}_{j}-\mathbf{r}_{l}|)=\frac{1}{V}\sum\limits_{\mathbf{k}}%
\nu(\mathbf{k})e^{i\mathbf{k}(\mathbf{r}_{j}-\mathbf{r}_{l})}, \label{11}%
\end{equation}
where $\mathbf{k}$ runs the values%
\begin{equation}
\mathbf{k}=2\pi\left(  \frac{j_{x}}{L_{x}},\frac{j_{y}}{L_{y}},\frac{j_{z}%
}{L_{z}}\right)  ; \label{12}%
\end{equation}
$j_{x}$, $j_{x}$, and $j_{x}$ are integers; $L_{x}$, $L_{y}$, and
$L_{z}$ are the system sizes; and  $V=L_{x}L_{y}L_{z}$.
Relations~(\ref{3}), (\ref{3p})  and (\ref{12}) ensure that the BCs
are satisfied.

Using the formulae $n_{0}=N_{0}/V$,%
\begin{equation}
a_{0}=N_{0}^{1/2}e^{i\theta},\quad a_{0}^{\ast}=N_{0}^{1/2}e^{-i\theta},
\label{5}%
\end{equation}%
\begin{equation}
a_{0}^{\ast}N_{0}^{-1/2}\hat{a}_{\mathbf{k}}=\frac{\hat{b}_{\mathbf{k}%
}+L_{\mathbf{k}}\hat{b}_{-\mathbf{k}}^{+}}{\sqrt{1-L_{\mathbf{k}}^{2}}},
\label{6}%
\end{equation}%
\begin{equation}
a_{0}N_{0}^{-1/2}\hat{a}_{\mathbf{k}}^{+}=\frac{\hat{b}_{\mathbf{k}}%
^{+}+L_{\mathbf{k}}\hat{b}_{-\mathbf{k}}}{\sqrt{1-L_{\mathbf{k}}^{2}}},
\label{6p}%
\end{equation}%
\begin{eqnarray}
L_{\mathbf{k}}&=&\frac{V}{N_{0}\nu(\mathbf{k})}\left[  E(k)-\frac{\hbar^{2}%
k^{2}}{2m}-n_{0}\nu(\mathbf{k})\right] \nonumber \\
&=&L_{-\mathbf{k}}=L_{|\mathbf{k}|},
\label{7}%
\end{eqnarray}%
\begin{equation}
E(k)=\sqrt{\left(  \frac{\hbar^{2}k^{2}}{2m}\right)  ^{2}+2n_{0}\nu
(k)\frac{\hbar^{2}k^{2}}{2m}}, \label{8}%
\end{equation}
where $\hat{a}_{\mathbf{k}}$ and $\hat{b}_{\mathbf{k}}$ are Bose operators,
Hamiltonian (\ref{2}) can be written in the diagonal form,%
\begin{equation}
\hat{H}_{\mathrm{app}}=E_{0}+\sum\limits_{\mathbf{k}\neq0}E(k)\hat
{b}_{\mathbf{k}}^{+}\hat{b}_{\mathbf{k}}. \label{9}%
\end{equation}
The formula for $E_{0}$ is written out in~\cite{bog1947}.

The monograph \cite{miransky1994} proposes the following
ground-state wave function (WF)  in the second quantisation
representation  (see
also the works~\cite{akhiezer1981,haldane1981}):%
\begin{eqnarray}
|\theta\rangle &=& e^{-N/2}e^{[N_{0}^{1/2}e^{i\theta}\hat{a}_{0}^{+}]}%
 \label{13}
\\ &\times&\prod\limits_{\mathbf{k}\neq0}(1-L_{|\mathbf{k}|}^{2})^{1/4}\exp{\left\{
\frac{1}{2}e^{2i\theta}L_{|\mathbf{k}|}\hat{a}_{\mathbf{k}}^{+}\hat
{a}_{-\mathbf{k}}^{+}\right\}  }|0_{\mathrm{bare}}\rangle,\nonumber%
\end{eqnarray}
where $|0_{\mathrm{bare}}\rangle$ is the vacuum state:%
\begin{equation}
\hat{a}_{\mathbf{k}}|0_{\mathrm{bare}}\rangle=0\quad \mbox{for all}%
\ \mathbf{k}.\label{14}%
\end{equation}
The state $|\theta\rangle$ generally describes an infinite system.
Since%
\begin{equation}
\hat{b}_{\mathbf{k}}|\theta\rangle\equiv\frac{\hat{a}_{\mathbf{k}}e^{-i\theta
}-L_{\mathbf{k}}e^{i\theta}\hat{a}_{-\mathbf{k}}^{+}}{\sqrt{1-L_{\mathbf{k}%
}^{2}}}|\theta\rangle=0\label{b1}%
\end{equation}
for all $\mathbf{k}\neq0$, the function $|\theta\rangle$ (\ref{13})
corresponds to the ground state:
$\hat{H}_{\mathrm{app}}|\theta\rangle =E_{0}|\theta\rangle$.

Let the WFs of the system transform according to the unitary law $\psi
_{n}\rightarrow\hat{U}\psi_{n}$, where $\hat{U}^{-1}=\hat{U}^{+}$. Then the
operators of physical quantities transform as $\hat{f}\rightarrow\hat{U}%
^{-1}\hat{f}\hat{U}$~\cite{land3}. Let%
\begin{equation}
\hat{U}_{\varphi}=e^{i\varphi\hat{N}},\quad\hat{N}=\sum\limits_{\mathbf{k}%
}\hat{a}_{\mathbf{k}}^{+}\hat{a}_{\mathbf{k}}.\label{2-u}%
\end{equation}
Then%
\begin{equation}
\hat{a}_{\mathbf{k}}\rightarrow\hat{U}_{\varphi}^{-1}\hat{a}_{\mathbf{k}}%
\hat{U}_{\varphi}=e^{i\varphi}\hat{a}_{\mathbf{k}},\label{16}%
\end{equation}
\begin{equation}
\hat{a}_{\mathbf{k}%
}^{+}\rightarrow\hat{U}_{\varphi}^{-1}\hat{a}_{\mathbf{k}}^{+}\hat{U}%
_{\varphi}=e^{-i\varphi}\hat{a}_{\mathbf{k}}^{+}\label{16p}%
\end{equation}
for all $\mathbf{k}$. This implies that%
\begin{equation}
\hat{\psi}(\mathbf{r},t)\rightarrow e^{i\varphi}\hat{\psi}(\mathbf{r}%
,t),\quad\hat{\psi}^{+}(\mathbf{r},t)\rightarrow e^{-i\varphi}\hat{\psi}%
^{+}(\mathbf{r},t).\label{15}%
\end{equation}
Note that $\hat{a}_{0}$ is an operator in formulae (\ref{16}) and
(\ref{15}), i.e. the replacement $\hat{a}_{0}\rightarrow a_{0}$ is
not made. Transformations (\ref{16}), (\ref{16p}) and (\ref{15})
define the $U(1)$ rotation. Using
the formulae $\hat{U}_{\varphi}|0_{\mathrm{bare}}\rangle=|0_{\mathrm{bare}%
}\rangle$, $\hat{U}_{\varphi}\exp{(\alpha\hat{a}_{0}^{+})}\hat{U}_{\varphi
}^{-1}=\exp{(\alpha\hat{U}_{\varphi}\hat{a}_{0}^{+}\hat{U}_{\varphi}^{-1}%
)}=\exp{(\alpha e^{i\varphi}\hat{a}_{0}^{+})}$, and $\hat{U}_{\varphi}%
\exp{(\beta\hat{a}_{\mathbf{k}}^{+}\hat{a}_{-\mathbf{k}}^{+})}\hat{U}%
_{\varphi}^{-1}=\exp{(\beta\hat{U}_{\varphi}\hat{a}_{\mathbf{k}}^{+}\hat
{U}_{\varphi}^{-1}\hat{U}_{\varphi}\hat{a}_{-\mathbf{k}}^{+}\hat{U}_{\varphi
}^{-1})}=\exp{(\beta e^{2i\varphi}\hat{a}_{\mathbf{k}}^{+}\hat{a}%
_{-\mathbf{k}}^{+})}$, it is easy to obtain%
\begin{eqnarray}
 \hat{U}_{\varphi}|\theta\rangle &=&
e^{-N/2}\exp{[N_{0}^{1/2}e^{i\theta}\hat{U}_{\varphi}\hat{a}^{+}_{0}\hat{U}^{-1}_{\varphi}]}\nonumber
\\ &\times& \prod\limits_{\mathbf{k}\neq
0}(1-L_{|\mathbf{k}|}^{2})^{1/4}e^{\hat{F}_{\mathbf{k}}}\hat{U}_{\varphi}|0_{\mathrm{bare}}\rangle
= |\theta+\varphi\rangle,
 \label{13-2} \end{eqnarray}
where
\begin{equation}
\hat{F}_{\mathbf{k}}=\frac{1}{2}e^{2i\theta}L_{|\mathbf{k}|}
\hat{U}_{\varphi}\hat{a}^{+}_{\mathbf{k}}\hat{U}^{-1}_{\varphi}
\hat{U}_{\varphi}\hat{a}^{+}_{-\mathbf{k}}\hat{U}^{-1}_{\varphi}.\label{13-2f}%
\end{equation}

The formulae above reproduce the results of the
work~\cite{miransky1994}. It is clear that the ground state
(\ref{13}) is not invariant under the $U(1)$ rotation
(\ref{2-u})--(\ref{15}). In this case the
exact Hamiltonian (\ref{1}) is invariant, $\hat{U}_{\varphi}^{-1}\hat{H}%
\hat{U}_{\varphi}=\hat{H}$, but the approximate Hamiltonian (\ref{2}) is not invariant:%
\begin{eqnarray}
&&
\hat{U}^{-1}_{\varphi}\hat{H}_{\mathrm{app}}\hat{U}_{\varphi}=\frac{N^{2}\nu(0)}{2V}
+ \sum\limits_{\mathbf{k}\neq 0}\frac{\hbar^{2}k^{2}}{2m}
\hat{a}^{+}_{\mathbf{k}}\hat{a}_{\mathbf{k}} \nonumber \\ && +
\sum\limits_{\mathbf{k}\neq 0}\frac{\nu(\mathbf{k})}{2V}\left
[a_{0}^{2}e^{-2i\varphi}\hat{a}^{+}_{\mathbf{k}}\hat{a}^{+}_{-\mathbf{k}}
+
(a_{0}^{*})^{2}e^{2i\varphi}\hat{a}_{\mathbf{k}}\hat{a}_{-\mathbf{k}}\right.
\nonumber \\ && + \left.
2a_{0}^{\ast}a_{0}\hat{a}^{+}_{\mathbf{k}}\hat{a}_{\mathbf{k}}
\right ] \neq \hat{H}_{\mathrm{app}}.
 \label{u2u} \end{eqnarray}
From Eq.~(\ref{u2u}), it follows that
$[\hat{H}_{\mathrm{app}},\hat{N}]\neq0$, i.e., the Bogoliubov
Hamiltonian (\ref{2}) does not conserve the number of particles.

The monograph \cite{miransky1994} suggests that the non-invariance
of the ground state (\ref{13}) with respect to the $U(1)$ rotation
indicates the spontaneous breakdown of the $U(1)$ symmetry.
Therefore, phonon excitations in an infinite, weakly interacting
Bose gas at near-zero temperatures are regarded as analogues to
Goldstone bosons. However, a closer examination reveals that the
situation is more subtle.
By definition, SSB is realised in an infinite system when the
Hamiltonian is invariant under the $U(1)$ symmetry, but the ground
state is not. In the present case, the situation is different: both
the ground state $|\theta\rangle$ and the Hamiltonian
$\hat{H}_{\mathrm{app}}$ are not invariant. In this case,
$|\theta\rangle$ is the ground state for the Hamiltonian $\hat{H}%
_{\mathrm{app}}$ rather than the exact Hamiltonian $\hat{H}$
(\ref{1}). One can show that the non-invariance of
$\hat{H}_{\mathrm{app}}$ always implies the
non-invariance of $|\theta\rangle$. To see this, assume by contradiction that $\hat{H}%
_{\mathrm{app}}$ is non-invariant, but $|\theta\rangle$ invariant,
i.e.,
$\hat{U}_{\varphi}^{-1}\hat{H}_{\mathrm{app}}\hat{U}_{\varphi}\equiv\hat
{H}_{\mathrm{app}}(\varphi)\neq\hat{H}_{\mathrm{app}}$ and
$\hat{U}_{\varphi }|\theta\rangle=e^{i\alpha}|\theta\rangle$
(strictly speaking, $|\theta \rangle$ is invariant when
$\hat{U}_{\varphi}|\theta\rangle=|\theta\rangle$; we write the more
general condition $\hat{U}_{\varphi}|\theta\rangle=e^{i\alpha
}|\theta\rangle$, since the factor $e^{i\alpha}$ can be included in
the
normalization constant of $|\theta\rangle$). Then $\hat{H}_{\mathrm{app}%
}|\theta\rangle=E_{0}|\theta\rangle$ and $\hat{U}_{\varphi}\hat{H}%
_{\mathrm{app}}|\theta\rangle=E_{0}\hat{U}_{\varphi}|\theta\rangle
=E_{0}e^{i\alpha}|\theta\rangle$. On the other hand,
$\hat{U}_{\varphi}\hat
{H}_{\mathrm{app}}|\theta\rangle=\hat{U}_{\varphi}\hat{H}_{\mathrm{app}}%
\hat{U}_{\varphi}^{-1}\hat{U}_{\varphi}|\theta\rangle=\hat{H}_{\mathrm{app}%
}(-\varphi)e^{i\alpha}|\theta\rangle\neq\hat{H}_{\mathrm{app}}e^{i\alpha
}|\theta\rangle=E_{0}e^{i\alpha}|\theta\rangle$. This contradiction
implies that $\hat{U}_{\varphi}|\theta\rangle\neq
e^{i\alpha}|\theta\rangle$. This means that the non-invariance of
the state $|\theta\rangle$ is a consequence of the non-invariance of
the Hamiltonian $\hat{H}_{\mathrm{app}}$, while the latter property
is related to the introduction of the $c$-number $a_{0}$.

Thus, at the level of the approximate Hamiltonian
$\hat{H}_{\mathrm{app}}$ and the WF $|\theta\rangle$,  SSB is
absent.
It remains unclear whether SSB occurs for the exact Hamiltonian
(\ref{1}) and corresponding to it exact ground state, since the
latter has not yet been found.
Perhaps the monograph \cite{miransky1994} implicitly assumes that
the exact and approximate ground states share the same symmetry
properties with respect to the $U(1)$ rotation, since these WFs are
expected to be close.
However, there is no reason to assume this: the Hamiltonians
(\ref{1}) and (\ref{2}) are also close, yet they possess different
symmetry properties with respect to the $U(1)$ rotation. This shows
that the Bogoliubov approach does not allow one to determine whether
SSB is present.

According to formulae (\ref{13}) and (\ref{13-2}), the ground state
of an infinite weakly interacting Bose gas is infinitely degenerate.
However, any real system is finite, and the number of particles in a
finite periodic system is fixed. Therefore, it is necessary to find
the ground-state WF, which is an eigenfunction of the particle
number operator $\hat{N}$  and corresponds to a finite
$N$. We did not find such a solution (WF~(\ref{13}) is not
an eigenfunction of the operator $\hat{N}$).

Note one more point. It follows from relation~(\ref{b1}) that $\hat
{a}_{\mathbf{k}\neq0}|\theta\rangle=L_{\mathbf{k}}e^{i2\theta}\hat
{a}_{-\mathbf{k}\neq0}^{+}|\theta\rangle$. Using this formula and
expanding both exponents in Eq.~(\ref{13}) into series, we see that
$\hat {a}_{\mathbf{k}\neq0}|\theta\rangle$ is a sum of terms
containing operators of the form $\hat{a}_{\mathbf{k}\neq 0}^{+}$
raised to odd powers only. On the other hand, the series expansion
of $|\theta\rangle$ contains the operators
$\hat{a}_{\mathbf{k}\neq0}^{+}$ raised to even powers only.
Therefore,
$\langle\theta|\hat{a}_{\mathbf{k}\neq0}|\theta\rangle=0$. Since $\hat{a}%
_{0}|\theta\rangle=N_{0}^{1/2}e^{i\theta}|\theta\rangle$, we obtain%
\begin{equation}
\langle\theta|\hat{\psi}(\mathbf{r},t)|\theta\rangle=V^{-1/2}\langle
\theta|\hat{a}_{0}|\theta\rangle=n_{0}^{1/2}e^{i\theta}. \label{ss}%
\end{equation}
This formula implies that $\hat{U}_{\varphi}|\theta\rangle\neq e^{i\alpha}%
|\theta\rangle$. Indeed, let $\hat{U}_{\varphi}|\theta\rangle=e^{i\alpha
}|\theta\rangle$. Then $\langle\theta|\hat{U}_{\varphi}^{-1}\hat{\psi
}(\mathbf{r},t)\hat{U}_{\varphi}|\theta\rangle=\langle\theta|\hat{\psi
}(\mathbf{r},t)|\theta\rangle$. On the other hand, according to Eqs.~(\ref{16}%
)--(\ref{15}), we obtain
{$\langle\theta|\hat{U}_{\varphi}^{-1}\hat{\psi
}(\mathbf{r},t)\hat{U}_{\varphi}|\theta\rangle=e^{i\varphi}\langle\theta
|\hat{\psi}(\mathbf{r},t)|\theta\rangle$}, which contradicts the
previous formula.
In view of this, the relation $\langle\theta|\hat{\psi
}(\mathbf{r},t)|\theta\rangle\neq 0$ implies that
$\hat{U}_{\varphi}|\theta\rangle\neq e^{i\alpha}%
|\theta\rangle$. Together with the formulae
$\hat{H}|\theta\rangle=E_{0}|\theta\rangle$ and
$\hat{U}^{-1}_{\varphi}\hat{H}\hat{U}_{\varphi} = \hat{H}$, this
leads to the conclusion that SSB is present and that the ground
state is degenerate.
Similar properties hold for quantum field
systems~\cite{miransky1994,goldstone1962,coleman}.

Thus, within the Bogoliubov approach, we could not derive the
ground-state WF for a finite system, nor did we ascertain whether a
spontaneous breakdown of $U(1)$ symmetry takes place in a finite or
infinite system.

\subsection{The particle-number-conserving Bogoliubov approach}

The reason for the failure in the previous section is the $c$-number
$\hat {a}_{0}=a_{0}=N_{0}^{1/2}e^{i\theta}$, which leads to the
non-invariance of $\hat{H}_{\mathrm{app}}$. Consider a more accurate
approach where the $c$-number is not used. In the
work~\cite{gardiner1997}, the Bogoliubov model was modified so that
the $c$-number was not introduced and the conservation law for the
particle number was satisfied. This line of approach was  developed
in the works~\cite{girardeau1998,zagrebnov2007}. The simplest
analysis was given in~\cite{zagrebnov2007}, where it was shown that
Bogoliubov's Hamiltonian can be written in the form%
\begin{eqnarray}
\hat{H}_{\mathrm{app,m}}&=&\frac{\hat{N}^{2}\nu(0)}{2V}-\frac{n\nu(0)}{2}%
+\sum\limits_{\mathbf{k}\neq0}\frac{\hbar^{2}k^{2}}{2m}\hat{\varsigma
}_{\mathbf{k}}^{+}\hat{\varsigma}_{\mathbf{k}}\nonumber
\\ &+&\frac{n}{2}\sum
\limits_{\mathbf{k}\neq0}\nu(k)\left[  \hat{\varsigma}_{\mathbf{k}}^{+}%
\hat{\varsigma}_{-\mathbf{k}}^{+}+\hat{\varsigma}_{\mathbf{k}}\hat{\varsigma
}_{-\mathbf{k}}+2\hat{\varsigma}_{\mathbf{k}}^{+}\hat{\varsigma}_{\mathbf{k}%
}\right]  , \label{2-1}%
\end{eqnarray}
where $n=N/V$, {
\begin{equation}
\hat{\varsigma}_{\mathbf{k}}=\hat{a}_{0}^{+}\left(
1+\hat{N}_{0}\right) ^{-1/2}\hat{a}_{\mathbf{k}},
\quad\mathbf{k}\neq 0, \label{2-2}%
\end{equation}
\begin{equation}
\hat{\varsigma}_{\mathbf{k}}^{+}=\hat
{a}_{\mathbf{k}}^{+}\left(  1+\hat{N}_{0}\right)  ^{-1/2}\hat{a}_{0}%
,\quad\mathbf{k}\neq 0, \label{2-2p}%
\end{equation}
}%
\begin{equation}
\lbrack\hat{\varsigma}_{\mathbf{k}},\hat{\varsigma}_{\mathbf{q}}%
]=[\hat{\varsigma}_{\mathbf{k}}^{+},\hat{\varsigma}_{\mathbf{q}}^{+}%
]=0,\quad\lbrack\hat{\varsigma}_{\mathbf{k}},\hat{\varsigma}_{\mathbf{q}}%
^{+}]=\delta_{\mathbf{k},\mathbf{q}},\ \ \mathbf{k},\mathbf{q}\neq0,
\label{2-c}%
\end{equation}
\begin{equation}
\hat{N}=\hat{N}_{0}+\sum\limits_{\mathbf{k}\neq 0}\hat{a}_{\mathbf{k}}^{+}%
\hat{a}_{\mathbf{k}}=\hat{N}_{0}+\sum\limits_{\mathbf{k}\neq0}\hat{\varsigma
}_{\mathbf{k}}^{+}\hat{\varsigma}_{\mathbf{k}}, \label{2-nn}%
\end{equation}
\begin{equation}
\hat{N}_{0}=\hat{a}_{0}^{+}\hat{a}_{0}, \label{2-n0}%
\end{equation}
and $(\hat{N}_{0}+1)^{\alpha}\hat{a}_{0}=\hat{a}_{0}\hat
{N}_{0}^{\alpha}$ for any real number $\alpha$~\cite{zagrebnov2007}.
The
operators $\hat{\varsigma}_{\mathbf{k}}$ and $\hat{\varsigma}_{\mathbf{k}}%
^{+}$ do not change the particle number:%
\begin{equation}
\lbrack\hat{N},\hat{\varsigma}_{\mathbf{k}}]=0,\quad\lbrack\hat{N}%
,\hat{\varsigma}_{\mathbf{k}}^{+}]=0\quad(\mathbf{k}\neq0). \label{2-n}%
\end{equation}
The Hamiltonian (\ref{2-1}) is similar to Bogoliubov's (\ref{2}) and
leads to  Bogoliubov's solutions for $E_{0}$ and $E(k)$. In this
case, the Hamiltonian (\ref{2-1}) preserves the number of particles,
$[\hat{H}_{\mathrm{app,m}},\hat{N}] = 0$.

In the work \cite{girardeau1998} a somewhat different Hamiltonian was obtained,%
\begin{eqnarray}
&&\hat{H}_{\mathrm{app,m2}}=\frac{\hat{N}(\hat{N}-1)\nu(0)}{2V} \nonumber\\
&&+ \sum\limits_{\mathbf{k}\neq 0}\left
(\frac{\hbar^{2}k^{2}}{2m}+\frac{\hat{N}_{0}\nu(\mathbf{k})}{V}\right
) \hat{a}^{+}_{\mathbf{k}}\hat{a}_{\mathbf{k}} \label{2-3}\\  &&+
\sum\limits_{\mathbf{k}\neq 0}\frac{\nu(\mathbf{k})}{2V}
[(\hat{N}_{0}+1)(\hat{N}_{0}+2)]^{1/2}\left
[\hat{\varsigma}^{+}_{\mathbf{k}}\hat{\varsigma}^{+}_{-\mathbf{k}} +
\hat{\varsigma}_{\mathbf{k}}\hat{\varsigma}_{-\mathbf{k}} \right
 ].
 \nonumber \end{eqnarray}
Taking Bogoliubov's approximations $N-N_{0}\ll N$ and $N\gg1$ into
account, this Hamiltonian is reduced to a simpler form~(\ref{2-1}).
In the paper \cite{gardiner1997}, a Hamiltonian was obtained for the
case of a point-like potential; it is equivalent to the Hamiltonian
(\ref{2-3}) if $N-N_{0}\ll N$ and $N\gg1$ (see
\cite{girardeau1998}).

We will rely on the Hamiltonian (\ref{2-1}). The terms
$\hat{N}^{2}\nu (0)/(2V)-n\nu(0)/2$ affect only the value of
$E_{0}$. The rest of the Hamiltonian is equivalent to  Bogoliubov's
Hamiltonian (\ref{2}) if we replace $a_{0}$ and $a_{0}^{\ast}$ by
$N^{1/2}$ in the latter. Instead of
formulae (\ref{6})--(\ref{8}), we have%
\begin{equation}
\hat{\varsigma}_{\mathbf{k}}=\frac{\hat{b}_{\mathbf{k}}+\tilde{L}_{\mathbf{k}%
}\hat{b}_{-\mathbf{k}}^{+}}{\sqrt{1-\tilde{L}_{\mathbf{k}}^{2}}},\quad
\hat{\varsigma}_{\mathbf{k}}^{+}=\frac{\hat{b}_{\mathbf{k}}^{+}+\tilde
{L}_{\mathbf{k}}\hat{b}_{-\mathbf{k}}}{\sqrt{1-\tilde{L}_{\mathbf{k}}^{2}}%
},d\mathbf{k}\neq0, \label{2-6}%
\end{equation}%
\begin{eqnarray}
\tilde{L}_{\mathbf{k}}&=&\frac{V}{N\nu(\mathbf{k})}\left[
E(k)-\frac{\hbar ^{2}k^{2}}{2m}-n\nu(\mathbf{k})\right] \nonumber
\\ &=&\tilde{L}_{-\mathbf{k}}=\tilde
{L}_{|\mathbf{k}|}, \label{2-7}%
\end{eqnarray}%
\begin{equation}
E(k)=\sqrt{\left(  \frac{\hbar^{2}k^{2}}{2m}\right)  ^{2}+2n\nu(k)\frac
{\hbar^{2}k^{2}}{2m}}, \label{2-8}%
\end{equation}
while formula (\ref{9}) does not change. As one can see, the phase
$\theta$ has dropped out of all formulae.

Given Eq.~(\ref{13}), it is easy to guess the ground-state WF for a finite
system of $N$ bosons:%
\begin{equation}
|0\rangle=\tilde{C}\exp{\left\{  \sum\limits_{\mathbf{k}\neq0}\frac{\tilde
{L}_{|\mathbf{k}|}}{2}\hat{\varsigma}_{\mathbf{k}}^{+}\hat{\varsigma
}_{-\mathbf{k}}^{+}\right\}  }[\hat{a}_{0}^{+}]^{N}|0_{\mathrm{bare}}%
\rangle.\label{2-9}%
\end{equation}
If $\nu(k)\rightarrow0$, then $\tilde{L}_{|\mathbf{k}|}\rightarrow0$ and
$|0\rangle\rightarrow\tilde{C}[\hat{a}_{0}^{+}]^{N}|0_{\mathrm{bare}}\rangle$.
For the WF~(\ref{2-9}) we get%
\begin{equation}
\hat{b}_{\mathbf{k}}|0\rangle=\frac{\hat{\varsigma}_{\mathbf{k}}-\tilde
{L}_{\mathbf{k}}\hat{\varsigma}_{-\mathbf{k}}^{+}}{\sqrt{1-\tilde
{L}_{\mathbf{k}}^{2}}}|0\rangle=0,\quad\hat{H}_{\mathrm{app}}|0\rangle
=E_{0}|0\rangle,\label{b2}%
\end{equation}%
\begin{eqnarray}
&&\hat{N}|0\rangle =  \hat{N}\tilde{C}\exp{\left
\{\sum\limits_{\mathbf{k}\neq 0}\frac{\tilde{L}_{|\mathbf{k}|}}{2}
\hat{\varsigma}^{+}_{\mathbf{k}}\hat{\varsigma}^{+}_{-\mathbf{k}}
\right \}}[\hat{a}^{+}_{0}]^{N}|0_{\mathrm{bare}}\rangle \nonumber\\
&&= \tilde{C}\exp{\left \{\sum\limits_{\mathbf{k}\neq
0}\frac{\tilde{L}_{|\mathbf{k}|}}{2}
\hat{\varsigma}^{+}_{\mathbf{k}}\hat{\varsigma}^{+}_{-\mathbf{k}}
\right \}}\hat{N}[\hat{a}^{+}_{0}]^{N}|0_{\mathrm{bare}}\rangle
=N|0\rangle,
 \label{2-10} \end{eqnarray}
\begin{equation}
\hat{U}_{\varphi}|0\rangle\equiv
e^{i\varphi\hat{N}}|0\rangle=e^{iN\varphi
}|0\rangle, \label{2-11}%
\end{equation}
\begin{equation}
\hat{U}_{\varphi}^{-1}\hat{H}_{\mathrm{app,m}}\hat{U}_{\varphi
}=\hat{H}_{\mathrm{app,m}}.\label{2-11b}%
\end{equation}
Formulae (\ref{2-11}), (\ref{2-11b}) show that the Hamiltonian and
the ground state are invariant under the $U(1)$
transformation~(\ref{16}). Periodic BCs for
$\hat{\psi}(\mathbf{r},t)$ are also invariant under the $U(1)$
rotation~(\ref{16}). Therefore, for a finite system of weakly
interacting bosons, the phonons are \textit{not} Goldstone bosons.
In addition, the formulae (\ref{b2}) and (\ref{2-10})
show that the function (\ref{2-9}) corresponds to the ground state
of a system with a total number of particles equal to $N$ (the
Hamiltonian (\ref{2-1}) was also obtained for $N$ particles because
the approximation $\hat{N}_{0}=N_{0}\approx N$~\cite{zagrebnov2007}
was used in its derivation).

The described properties are consistent with the Noether theorem,
according to which the invariance of the action (Lagrangian,
Hamiltonian) under transformations of a continuous symmetry group
leads to a conservation law for some \textquotedblleft
charge\textquotedblright. In our case this is the group $U(1)$ and
the conservation law for the number of particles.

Thus, the approach based on the number-conserving Hamiltonian
(\ref{2-1}) allows one to find the ground state of a finite system
of $N$ spinless bosons and to establish that there is no spontaneous
breaking of $U(1)$ symmetry in such a system.

\subsection{The approach based on exact wave functions}

Although the modified Bogoliubov approach considered above is more
accurate than the standard Bogoliubov approach, it still remains
approximate. However, the question posed in the title of this
article can be answered on the basis of \textit{exact} formulae.

The exact ground-state wave function of a periodic system of $N$
interacting spinless
bosons, which takes into account two-particle and higher-order correlations, reads%
\begin{eqnarray}
\ln{\Psi_{0}} &=& \ln{C}+\sum\limits_{j_{1},j_{2}=1}^{N}
S_{2}(\mathbf{r}_{j_{1}j_{2}})\nonumber\\ &+&
\sum\limits_{j_{1},j_{2},j_{3}=1}^{N}
S_3(\mathbf{r}_{j_{1}j_{2}},\mathbf{r}_{j_{2}j_{3}},\mathbf{r}_{j_{3}j_{1}})
+ \ldots \nonumber\\ &+&  \sum\limits_{j_{1},\ldots,j_{N}=1}^{N}
S_{N}(\mathbf{r}_{j_{1}j_{2}},\mathbf{r}_{j_{2}j_{3}},\ldots,\mathbf{r}_{j_{N}j_{1}}),
     \label{3-1} \end{eqnarray}
where $\mathbf{r}_{lj}=\mathbf{r}_{l}-\mathbf{r}_{j}$. The WF
(\ref{3-1}) describes the ground state of a Bose system at any
coupling strength: weak, intermediate, or strong. In other words, it
describes both the Bose gas and the Bose liquid
\cite{yuv1,holes2020,gross1962,woo1972,feenberg1974}, as well as the
Bose crystal
\cite{mcmillan1965,chester1970,reatto1995,whitlock2006,mt2022}.
Using
the collective variables $\rho_{\mathbf{k}}=N^{-1/2}\sum_{j=1}%
^{N}e^{-i\mathbf{k}\mathbf{r}_{j}}$, formula~(\ref{3-1}) can be written in the
form \cite{yuv1} (see also \cite{holes2020}):%
\begin{equation}
\Psi_{0}(\mathbf{r}_{1},\ldots,\mathbf{r}_{N})=A_{0}e^{S(\mathbf{r}_{1}%
,\ldots,\mathbf{r}_{N})}, \label{3-2-1}%
\end{equation}
\begin{eqnarray}
S&=& \sum\limits_{\mathbf{q}_{1}\neq
0}\frac{c_{2}(\mathbf{q}_{1})}{2!}\rho_{\mathbf{q}_{1}}
\rho_{-\mathbf{q}_{1}}\nonumber\\  &+&
\sum\limits_{\mathbf{q}_{1},\mathbf{q}_{2}\neq
0}^{\mathbf{q}_{1}+\mathbf{q}_{2}\not= 0}
\frac{c_{3}(\mathbf{q}_{1},\mathbf{q}_{2})}{3!N^{1/2}}
\rho_{\mathbf{q}_{1}}\rho_{\mathbf{q}_{2}}\rho_{-\mathbf{q}_{1} -
\mathbf{q}_{2}}+\ldots + \nonumber\\  &+&
\sum\limits_{\mathbf{q}_{1},\ldots,\mathbf{q}_{N-1}\neq
0}^{\mathbf{q}_{1}+\ldots +\mathbf{q}_{N-1}\not= 0}
\frac{c_{N}(\mathbf{q}_{1},\ldots,\mathbf{q}_{N-1})}{N!N^{(N-2)/2}}
\nonumber\\
&\times&\rho_{\mathbf{q}_1}\ldots\rho_{\mathbf{q}_{N-1}}
\rho_{-\mathbf{q}_{1} - \ldots - \mathbf{q}_{N-1}}.
    \label{3-3-1}   \end{eqnarray}

To ascertain the properties of the ground state with respect to the
$U(1)$ transformations (\ref{16}), (\ref{16p}), let us express
$\rho_{\mathbf{k}}$ in terms of the particle creation and
annihilation operators, $\hat{a}_{\mathbf{q}}^{+}$ and
$\hat{a}_{\mathbf{q}}$. It follows from the formulae
$\hat{\rho}(\mathbf{r})=\hat{\psi
}^{+}(\mathbf{r})\hat{\psi}(\mathbf{r})$ and (\ref{3}), (\ref{3p})
that {\large
\begin{equation}
\hat{\rho}(k)=\int dr\hat{\rho}(r)e^{-i\mathbf{k}\mathbf{r}}=\sum_{\mathbf{q}%
}\hat{a}_{\mathbf{q}}^{+}\hat{a}_{\mathbf{q}+\mathbf{k}}. \label{3-4-1}%
\end{equation}
}On the other hand, $\hat{\rho}(\mathbf{r})=\sum_{j=1}^{N}\delta
(\mathbf{r}-\mathbf{r}_{j})$, so that
\begin{equation}
\hat{\rho}(\mathbf{k})=\int d\mathbf{r}\hat{\rho}(\mathbf{r})e^{-i\mathbf{k}%
\mathbf{r}}=\sum\limits_{j=1}^{N}e^{-i\mathbf{k}\mathbf{r}_{j}}. \label{3-4-2}%
\end{equation}
This gives us the desired formula \cite{bp1953,pn1}:
\begin{eqnarray}
\rho_{\mathbf{k}\neq0}&=&\frac{\hat{\rho}(\mathbf{k}\neq0)}{\sqrt{N}}=\frac
{1}{\sqrt{N}}\sum_{j=1}^{N}e^{-i\mathbf{k}\mathbf{r}_{j}}\nonumber \\ &=&\frac{1}{\sqrt{N}%
}\sum_{\mathbf{q}}\hat{a}_{\mathbf{q}-\mathbf{k}}^{+}\hat{a}_{\mathbf{q}}.
\label{3-4-3}%
\end{eqnarray}

If the interatomic interaction tends to zero, then
$c_{j\geq2}\rightarrow0$ in Eq.~(\ref{3-3-1}) \cite{yuv1}, and the
ground-state WF $|0\rangle$ of $N$ interacting bosons must reduce to
the WF of $N$ free bosons,
$(N!)^{-1/2}[\hat{a}_{0}^{+}]^{N}|0_{\mathrm{bare}}\rangle$. This
property---together with formulae (\ref{3-2-1}), (\ref{3-3-1}), and
(\ref{3-4-3})---makes it possible to write down the exact
ground-state wave function in terms of operators
$\hat{a}_{\mathbf{q}}^{+}$ and $\hat
{a}_{\mathbf{q}}$:%

\begin{equation}
|0\rangle=A_{0}e^{\hat{S}}[\hat{a}_{0}^{+}]^{N}|0_{\mathrm{bare}}%
\rangle,\label{3-2-2}%
\end{equation}
\begin{eqnarray}
\hat{S}&=& \sum\limits_{\mathbf{q}_{1}\neq
0}\frac{c_{2}(\mathbf{q}_{1})}{2!}\hat{\rho}_{\mathbf{q}_{1}}
\hat{\rho}_{-\mathbf{q}_{1}}\nonumber\\  &+&
\sum\limits_{\mathbf{q}_{1},\mathbf{q}_{2}\neq
0}^{\mathbf{q}_{1}+\mathbf{q}_{2}\not= 0}
\frac{c_{3}(\mathbf{q}_{1},\mathbf{q}_{2})}{3!N^{1/2}}
\hat{\rho}_{\mathbf{q}_{1}}\hat{\rho}_{\mathbf{q}_{2}}\hat{\rho}_{-\mathbf{q}_{1}
- \mathbf{q}_{2}}+\ldots + \nonumber\\  &+&
\sum\limits_{\mathbf{q}_{1},\ldots,\mathbf{q}_{N-1}\neq
0}^{\mathbf{q}_{1}+\ldots +\mathbf{q}_{N-1}\not= 0}
\frac{c_{N}(\mathbf{q}_{1},\ldots,\mathbf{q}_{N-1})}{N!N^{(N-2)/2}}
\nonumber\\  &\times&
\hat{\rho}_{\mathbf{q}_1}\ldots\hat{\rho}_{\mathbf{q}_{N-1}}
 \hat{\rho}_{-\mathbf{q}_{1} - \ldots - \mathbf{q}_{N-1}},
    \label{3-3-2}   \end{eqnarray}
where we denote%
\begin{equation}
\hat{\rho}_{\mathbf{k}\neq0}=\frac{1}{\sqrt{N}}\sum_{\mathbf{q}}\hat
{a}_{\mathbf{q}-\mathbf{k}}^{+}\hat{a}_{\mathbf{q}}.\label{3-4-4}%
\end{equation}
The operator $\hat{\rho}_{\mathbf{k}}$~(\ref{3-4-4}) is
invariant under the $U(1)$ rotation,%
\begin{eqnarray}
\hat{U}_{\varphi}^{-1}\hat{\rho}_{\mathbf{k}\neq0}\hat{U}_{\varphi}&=&\frac
{1}{\sqrt{N}}\sum_{\mathbf{q}}\hat{U}_{\varphi}^{-1}\hat{a}_{\mathbf{q}%
-\mathbf{k}}^{+}\hat{U}_{\varphi}\hat{U}_{\varphi}^{-1}\hat{a}_{\mathbf{q}%
}\hat{U}_{\varphi}\nonumber \\ &=&\hat{\rho}_{\mathbf{k}\neq0}.\label{3-ro}%
\end{eqnarray}
Therefore, the ground state (\ref{3-2-2}), (\ref{3-3-2}) is also
invariant,
\begin{eqnarray}
\hat{U}_{\varphi}|0\rangle
&=&A_{0}\hat{U}_{\varphi}e^{\hat{S}}\hat{U}_{\varphi
}^{-1}\hat{U}_{\varphi}[\hat{a}_{0}^{+}]^{N}|0_{\mathrm{bare}}\rangle
\nonumber \\ &=&A_{0}e^{\hat{S}}\hat{U}_{\varphi}[\hat{a}_{0}^{+}]^{N}|0_{\mathrm{bare}%
}\rangle=e^{i\varphi N}|0\rangle.\label{3-gs}%
\end{eqnarray}
To obtain Eq.~(\ref{3-gs}), one can expand $e^{\hat{S}}$ into a
series and
use formulae (\ref{3-3-2}), (\ref{3-ro}), and the relation%
\begin{eqnarray}
&&\hat{U}_{\varphi}\hat{\rho}_{\mathbf{q}_{1}}\ldots\hat{\rho}_{\mathbf{q}%
_{N-1}}\hat{U}_{\varphi}^{-1}\nonumber \\&&=\hat{U}_{\varphi}\hat{\rho}_{\mathbf{q}_{1}}%
\hat{U}_{\varphi}^{-1}\hat{U}_{\varphi}\hat{\rho}_{\mathbf{q}_{2}}\ldots
\hat{U}_{\varphi}^{-1}\hat{U}_{\varphi}\hat{\rho}_{\mathbf{q}_{N-1}}\hat
{U}_{\varphi}^{-1}\nonumber \\&&=\hat{\rho}_{\mathbf{q}_{1}}\ldots\hat{\rho}_{\mathbf{q}%
_{N-1}}.\label{3-ro2}%
\end{eqnarray}

Similarly, any excited state with the momentum $\mathbf{p}$ is described by
the WF \cite{yuv2,holes2020}%
\begin{equation}
|\mathbf{p}\rangle=A_{\mathbf{p}}\hat{\psi}_{\mathbf{p}}|0\rangle,
\label{3-e1}%
\end{equation}
where%
       \begin{eqnarray}
\hat{\psi}_{\mathbf{p}} & =&
b_{1}(\mathbf{p})\hat{\rho}_{-\mathbf{p}} +
\sum\limits_{\mathbf{q}_{1}\neq 0}^{\mathbf{q}_{1}+\mathbf{p}\neq 0}
\frac{b_{2}(\mathbf{q}_{1};\mathbf{p})}{2!N^{1/2}}
\hat{\rho}_{\mathbf{q}_{1}}\hat{\rho}_{-\mathbf{q}_{1}-\mathbf{p}}
\nonumber \\ &+& \sum\limits_{\mathbf{q}_{1},\mathbf{q}_{2}\neq
0}^{\mathbf{q}_{1}+ \mathbf{q}_{2}+\mathbf{p} \not= 0}
\frac{b_{3}(\mathbf{q}_{1},\mathbf{q}_{2};\mathbf{p})}{3!N}
\hat{\rho}_{\mathbf{q}_{1}}\hat{\rho}_{\mathbf{q}_{2}}\hat{\rho}_{-\mathbf{q}_{1}-\mathbf{q}_{2}-\mathbf{p}}
 \nonumber \\ &+& \ldots +
\sum\limits_{\mathbf{q}_{1},\ldots,\mathbf{q}_{N-1}\neq
0}^{\mathbf{q}_{1}+ \ldots +\mathbf{q}_{N-1}+\mathbf{p}\not= 0}
\frac{b_{N}(\mathbf{q}_{1},\ldots,\mathbf{q}_{N-1};\mathbf{p})}{N!N^{(N-1)/2}}
\nonumber \\
&\times&\hat{\rho}_{\mathbf{q}_1}\ldots\hat{\rho}_{\mathbf{q}_{N-1}}
\hat{\rho}_{-\mathbf{q}_{1} - \ldots - \mathbf{q}_{N-1}-\mathbf{p}},
       \label{3-e2}\end{eqnarray}
and the state $|0\rangle$ is given by formulae (\ref{3-2-2}) and
(\ref{3-3-2}). The order of operators $\hat{\rho}_{\mathbf{q}}$ in
formulae (\ref{3-3-2}) and (\ref{3-e2}) does not matter because
$[\hat{\rho }_{\mathbf{q}_{1}},\hat{\rho}_{\mathbf{q}_{2}}]=0$.
Formulae (\ref{3-e1}) and (\ref{3-e2}) are exact. From
Eqs.~(\ref{3-ro}), (\ref{3-ro2}), and (\ref{3-e2}), it follows that
$\hat{U}_{\varphi}\hat{\psi}_{\mathbf{p}}\hat
{U}_{\varphi}^{-1}=\hat{\psi}_{\mathbf{p}}$. Therefore, the function
$|\mathbf{p}\rangle$ (\ref{3-e1}) is invariant with respect to the
$U(1)$ transformations (\ref{16}), (\ref{16p}):
$\hat{U}_{\varphi}|\mathbf{p}\rangle
=A_{\mathbf{p}}\hat{U}_{\varphi}\hat{\psi}_{\mathbf{p}}\hat{U}_{\varphi}%
^{-1}\hat{U}_{\varphi}|0\rangle=e^{i\varphi N}|\mathbf{p}\rangle$.

Let us show that this inference does not change for the degenerate
ground state. Degeneracy means that the ground state corresponds to
several different WFs, one of which has no nodes, while the others
have nodes. Each state with nodes can be described as an excited
state (\ref{3-e1}), for which the total energy and the total
momentum of the excitations are zero ($\mathbf{p}=0$ in
Eqs.~(\ref{3-e1}) and (\ref{3-e2}), with $\hat{\rho
}_{\mathbf{p}=0}=N^{1/2}$). Since the function $|\mathbf{p}\rangle$
(\ref{3-e1}) is invariant under the $U(1)$ rotation (\ref{16}),
(\ref{16p}), this degeneracy of the ground state is not related to
the breaking of the $U(1)$ symmetry.

Hence, \emph{a spontaneous breaking of $U(1)$ symmetry is absent in
a finite system of $N$ interacting spinless bosons}. This is a
general conclusion that applies to a Bose gas, a Bose liquid, and a
Bose crystal.

Note that both Eq.~(\ref{2-9}) and Eqs.~(\ref{3-2-2})--(\ref{3-4-4}) lead to
the equality%
\begin{equation}
\langle0|\hat{\psi}(\mathbf{r},t)|0\rangle=0 \label{3-psi}%
\end{equation}
because the states $\hat{\psi}(\mathbf{r},t)|0\rangle$ and $\langle0|$
describe systems with $N-1$ and $N$ particles, respectively.

In the case of a weakly nonideal Bose gas, we have $c_{3}\approx c_{4}%
\approx\ldots\approx c_{N}\approx0$ \cite{bz,yuv1}, and formulae
(\ref{3-2-2})
and (\ref{3-3-2}) take the form%
\[
|0\rangle=Ce^{\hat{S}_{2}}[\hat{a}_{0}^{+}]^{N}|0_{\mathrm{bare}}\rangle
,\quad\hat{S}_{2}=\frac{1}{2}\sum\limits_{\mathbf{k}\neq0}c_{2}(\mathbf{k}%
)\hat{\rho}_{\mathbf{k}}\hat{\rho}_{-\mathbf{k}}.
\]
The equation $\hat{H}|0\rangle=E_{0}|0\rangle$ should lead to the
known solutions \cite{bz,yuv1} for $E_{0}$ and $c_{2}(\mathbf{k})$.
However, as far as we
know, $E_{0}$ and $c_{2}%
(\mathbf{k})$ have not yet been found  within this approach.

According to group theory, if the accidental degeneracy is absent,
then the degeneracy multiplicity of a state equals the dimension of
the irreducible representation (of the symmetry group of the
boundary value problem) according to which the WF of that state
transforms (see Appendix). We consider an Abelian compact group
$U(1)$ for which all irreducible representations are unitary and
one-dimensional: $T^{(l)}=e^{il\varphi}$,%
$l=0,\pm1,\pm2,\ldots$~\cite{elliott}. In this case, the operators
$\hat{T}(\varphi)=e^{i\varphi\hat{N}}$ form a group that is
isomorphic to $U(1)$, and $\hat{N}$ is the generator of this group
of operators. For any state $|\mathbf{p}\rangle$ of a system of $N$
bosons,
including the ground state $|0\rangle$, we obtained above $e^{i\varphi\hat{N}%
}|\mathbf{p}\rangle=e^{iN\varphi}|\mathbf{p}\rangle$. This fact
implies that each such state  transforms according to the same
one-dimensional representation $e^{iN\varphi}$ of the $U(1)$ group
and is therefore non-degenerate with respect to this group. This is
not surprising because the representation $e^{iN\varphi}$
corresponds to the quantum number $N$, which is the same for all
considered states $|\mathbf{p}\rangle$. In this
case, any excited state of the system is degenerate, $E(\mathbf{p}%
)=E(|\mathbf{p}|)$, because the momentum and inversion operators commute with
$\hat{H}$, but not with each other~\cite{sgs}.

Formulae (\ref{3-2-2}) and (\ref{3-3-2}) specify  the exact
many-particle ground-state WF of the system, written in the
single-particle language in the second quantisation representation.
This \textquotedblleft single-particle\textquotedblright\ approach
is much more complicated than those based on the language of
elementary quasiparticles: for example, calculating the ground-state
energy within the single-particle approach would be exceedingly
laborious. However, formulae (\ref{3-2-2}) and (\ref{3-3-2}) allow
one to precisely determine whether spontaneous breaking of $U(1)$
symmetry occurs.

\section{Origin of the degeneracy of the ground state of an infinite Bose gas,
the method of quasi-averages, and the $1/q^{2}$-theorem}

Referring to Bogoliubov's $1/q^{2}$-theorem
\cite{bogquasi1,bogquasi2}, many authors of monographs and articles
have claimed that the ground state of a Bose gas is degenerate.
However, in section~2 we showed that the ground state of a
\textit{finite} Bose gas is not degenerate. Let us find out the
origin of this discrepancy.

The method of quasi-averages and the $1/q^{2}$-theorem were first
proposed in the preprint~\cite{bogquasi1} and later published in the
monograph~\cite{bogquasi2}. The method is based on introducing a
small term $\delta\hat{H}=\nu f(\hat {a}_{0},\hat{a}_{0}^{+})$ to
the standard Hamiltonian of the Bose gas, thereby violating the
$U(1)$ symmetry of the Hamiltonian. Bogoliubov  employed the
principle of attenuation of correlations \cite{bogquasi1,bogquasi2}{
\begin{equation}
\langle\hat{A}(r_{1})\hat{B}(r_{2})\rangle_{q}|_{|\mathbf{r}_{1}%
-\mathbf{r}_{2}|\rightarrow\infty}\rightarrow\langle\hat{A}(r_{1})\rangle
_{q}\cdot\langle\hat{B}(r_{2})\rangle_{q}, \label{4-1}%
\end{equation}
}where  $\langle\rangle_{q}$ denotes the quasi-average,
$\langle\hat{A}(\mathbf{r})\rangle_{q}=\lim\limits_{\nu\rightarrow0}%
(\lim\limits_{N,V\rightarrow\infty}\langle\hat{A}(\mathbf{r})\rangle)$,
and $\langle\rangle$ is the usual statistical average. The averages
$\langle \rangle_{q}$ and $\langle\rangle$ are found for the
Hamiltonian with and without the term $\delta\hat{H}$, respectively.
It has been postulated~\cite{bogquasi1,bogquasi2} that almost all
the atoms of a weakly interacting Bose gas are in the condensate at
low temperatures. Applying
Eq.~(\ref{4-1}) and putting $\hat{A}(\mathbf{r}_{1})=\hat{\psi}^{+}%
(\mathbf{r}_{1})$ and
$\hat{B}(\mathbf{r}_{2})=\hat{\psi}(\mathbf{r}_{2})$, one obtains
the condensate number density $n_{0}$ on the left-hand side of
Eq.~(\ref{4-1}). Then the right-hand side of Eq.~(\ref{4-1}) yields
$\langle\hat{\psi}(\mathbf{r})\rangle_{q}=e^{i\varphi}\sqrt{n_{0}}$
and
$\langle\hat{\psi}^{+}(\mathbf{r})\rangle_{q}=e^{-i\varphi}\sqrt{n_{0}}$,
although $\langle\hat{\psi}(\mathbf{r})\rangle=0$. Here the phase
$\varphi$ is arbitrary. Bogoliubov  used the term
$\delta\hat{H}=-\nu\sqrt{V}(\hat {a}_{0}+\hat{a}_{0}^{+})$, which
led to the phase choice $\varphi=0$. These relations led to the
introduction of the $c$-number $\hat{a}_{0}=\sqrt{N_{0}}$, and the
analysis resulted in the Bogoliubov dispersion law of quasiparticles
\cite{bogquasi1,bogquasi2}.

The method of quasi-averages is valid for $T>0$ alone. However, we
can let $T$ go to zero, $T\rightarrow0$, then the statistical
average becomes a
quantum-mechanical average: $\langle\hat{\psi}(\mathbf{r})\rangle_{q}%
|_{T\rightarrow0}\rightarrow\langle0|\hat{\psi}(\mathbf{r})|0\rangle_{q}$.
It is intuitively clear that at very low but nonzero temperatures,
statistical degeneracy  implies ground-state degeneracy. We accept
this here without proof. In view of this property, we consider only
the ground state ($T= 0$), although Bogoliubov considered
statistical averages ($T> 0$).

According to Bogoliubov's idea, if an arbitrarily small
$\delta\hat{H}(\varphi)$ leads to a non-negligible quasi-average
$\langle\hat{\psi}(\mathbf{r})\rangle_{q}=e^{i\varphi}\sqrt{n_{0}}$,
we have an infinitely degenerate ground state $|0\rangle$ and
SSB~\cite{bogquasi1,bogquasi2}. The degeneracy is related to the
fact that $\delta\hat{H}(\varphi)$ changes the system energy $E_{0}$
by an infinitely small value, so all $E_{0}(\varphi)$ can be
considered as identical. The SSB arises because the original
Hamiltonian $\hat{H}$ is invariant with respect to the $U(1)$
rotation, but the ground state is not (the inequality
$\langle\theta|\hat{\psi}(\mathbf{r},t)|\theta\rangle\neq0$ means
that $\hat{U}_{\varphi}|\theta\rangle\neq
e^{i\alpha}|\theta\rangle$, in this case
$\hat{H}\hat{U}_{\varphi}|\theta\rangle=\hat{U}_{\varphi}\hat
{H}|\theta\rangle=E_{0}\hat{U}_{\varphi}|\theta\rangle$ inasmuch as $[\hat{H}%
,\hat{N}]=0$). More precisely, Bogoliubov derived the inequality
$\langle\hat{\psi}(\mathbf{r},t)\rangle_{q} \neq 0$ instead of
$\langle\theta|\hat{\psi}(\mathbf{r},t)|\theta\rangle\neq0$.
However, he believed that the introduction of
$\delta\hat{H}(\varphi)$ merely reveals the degeneracy inherent in
the unperturbed statistical equilibrium state, and that the usual
average is zero ($\langle\hat{\psi}(\mathbf{r})\rangle=0$) because
it contains the averaging over $\varphi$. Since the non-zero
quasi-average
$\langle\hat{\psi}(\mathbf{r})\rangle_{q}=\sqrt{n_{0}}$ leads to a
gapless dispersion law, the $1/q^{2}$-theorem is similar to the
Goldstone theorem.

Thus, the $1/q^{2}$-theorem states that SSB can occur in the Bose
gas, which contradicts our results obtained in section 2. This
contradiction arose because we considered a finite system, whereas
Bogoliubov an infinite one. For a finite system, the quasi-average
transforms into the ordinary average:
\[
\langle\hat{\psi}(\mathbf{r})\rangle_{q}\equiv\lim\limits_{\nu\rightarrow
0}\langle\hat{\psi}(\mathbf{r})\rangle=\langle\hat{\psi}(\mathbf{r}%
)\rangle=0.
\]
However, for an infinite system, it is possible that
\[
\langle\hat{\psi}(\mathbf{r})\rangle_{q}\equiv\lim\limits_{\nu\rightarrow
0}(\lim\limits_{N,V\rightarrow\infty}\langle\hat{\psi}(\mathbf{r})\rangle
)\neq\lim\limits_{N,V\rightarrow\infty}\lim\limits_{\nu\rightarrow0}%
\langle\hat{\psi}(\mathbf{r})\rangle=0
\]
because the limits $\nu\rightarrow0$ and $ N,V\rightarrow\infty$ may
not commute.

It is important to understand the nature of the ground-state
degeneracy in an infinite Bose gas. Consider a periodic system of
$N$ \textit{free} spinless
bosons. If $N$ is finite, then the ground-state WF is%
\begin{equation}
\Psi_{0}(\mathbf{r}_{1},\ldots,\mathbf{r}_{N})=\left(  \frac{1}{\sqrt{V}%
}\right)  ^{N}. \label{4-3}%
\end{equation}
This state is non-degenerate, and all the atoms are in the
condensate $\psi(\mathbf{r})=V^{-1/2}$. Since any wave function is
determined up to the factor $e^{i\beta}$, the wave function
$\Psi_{0}$ (\ref{4-3}) can be written in the equivalent form
\begin{equation}
\Psi_{0}(\mathbf{r}_{1},\ldots,\mathbf{r}_{N})=e^{iN\alpha}\left(  \frac
{1}{\sqrt{V}}\right)  ^{N}=\left(  \frac{e^{i\alpha}}{\sqrt{V}}\right)  ^{N}.
\label{4-4}%
\end{equation}
Then the condensate WF is $\psi(\mathbf{r})=e^{i\alpha}V^{-1/2}$.
For any phase $\alpha$, function (\ref{4-4}) corresponds to the same
ground state (\ref{4-3}). This means the phase degeneracy, but such
a degeneracy is fictitious in this case.

If $N$ is infinite, the picture is more interesting. Let us pass to
an infinite system using a standard technique of statistical physics
--- the thermodynamic limit $ N,V \rightarrow\infty$ with
$N/V=\mathrm{const}$. Then the ground state WF of an infinite system
of
free bosons reads%
\begin{equation}
\Psi_{0}(\mathbf{r}_{1},\ldots,\mathbf{r}_{N})|_{N\rightarrow\infty}%
=\lim\limits_{N,V\rightarrow\infty}\left(  \frac{e^{i\alpha}}{\sqrt{V}%
}\right)  ^{N}, \label{4-5}%
\end{equation}
where $N/V=n=\mathrm{const}$. For a system of $N+1$ particles in the volume
$\acute{V}=(N+1)V/N$,%
\begin{equation}
\Psi_{0}(\mathbf{r}_{1},\ldots,\mathbf{r}_{N},\mathbf{r}_{N+1})|_{N\rightarrow
\infty}=\lim\limits_{N,\acute{V}\rightarrow\infty}\left(  \frac{e^{i\alpha}%
}{\sqrt{\acute{V}}}\right)  ^{N+1}. \label{4-6}%
\end{equation}
Let us exploit the fact that, when $N$ is infinite,  $\sqrt{\acute{V}}^{N+1}%
=\sqrt{V}^{N}$ because $N+1=N$. Then Eq.~(\ref{4-6}) can be
rewritten as
follows:%
\begin{equation}
\Psi_{0}(\mathbf{r}_{1},\ldots,\mathbf{r}_{N})|_{N\rightarrow\infty}%
=\lim\limits_{N,V\rightarrow\infty}\left(  \frac{e^{i(\alpha+\delta\alpha)}%
}{\sqrt{V}}\right)  ^{N}, \label{4-7}%
\end{equation}
where the phase $\alpha\in]0.2\pi\lbrack$ acquired the increment
$\delta \alpha=\alpha/N\rightarrow0$. If we similarly consider
systems of $N+j$ particles with $j=2,3,\ldots,\lfloor2\pi
N/\alpha\rfloor$, we obtain the WF (\ref{4-7}) and the condensate
$\psi(\mathbf{r})=e^{i(\alpha+\delta\alpha)}V^{-1/2}$, where the
phase $\alpha+\delta\alpha$ takes all possible values in the
interval $[\alpha ,\alpha+2\pi]$.

Thus, for an infinite system with $N=\infty$, the phase degeneracy
can be obtained, because adding particles to such a system does not
change the total number of particles: $\infty+j=\infty$ for all
$j=1,2,\ldots,\infty $. In this case, the ground state is infinitely
degenerate, since $E_{0}(N+j)=E_{0}(N)$. On the other hand, infinite
systems with different $N$ but the same particle number density $n$
are indistinguishable (in particular, all functions (\ref{4-7}) are
nodeless; however, in the case of ordinary quantum-mechanical
degeneracy, the WFs have various nodal structures and, in principle,
they are experimentally distinguishable). Therefore, we can treat
such systems as the same system. Then the ground state is
non-degenerate. Hence, the ground state of such an infinite system
can be regarded as both non-degenerate and infinitely degenerate.
This property can be added to the numerous paradoxes \cite{kline}
that arise from infinity.

For an infinite system of \textit{interacting} bosons, the phase
degeneracy of the WF (\ref{13}) and of the condensate
\cite{bogquasi1,bogquasi2} is also related to the indeterminacy of
the number of particles, $N$. This is evident from the following. In
sections 2.2 and 2.3, we found that the ground-state WF of a finite
system of $N$ interacting bosons transforms according to the
one-dimensional representation $e^{iN\varphi}$ of the $U(1)$ group,
so that such a state is non-degenerate with respect to this group.
However, the ground state of an infinite system is infinitely
degenerate with respect to this group (according to formulae
(\ref{13}) and (\ref{13-2})). Since all irreducible representations
of the $U(1)$ group are one-dimensional, we have an accidental
degeneracy: an infinite number of different representations $e^{i
N\varphi}$ correspond to the same energy. This is really the case
because for $N=\infty$ we have $N\pm j=N$, so that
$e^{i(N\pm1)\varphi}=e^{i(N\pm2)\varphi}=\ldots=e^{iN\varphi}$ and
$E_{0}(N\pm1)=E_{0}(N\pm2)=\ldots=E_{0}(N)$.
This shows that the ground-state degeneracy of an infinite system of
interacting bosons occurs precisely because $N$ is indefinite at
$N=\infty$. Similarly to an ideal gas, the ground state can be
considered simultaneously as both non-degenerate and infinitely
degenerate.

Bogoliubov supposed that the source of statistical degeneracy for an
infinite system of spinless bosons (interacting or free) is
different, namely, this is the conservation law for the number of
particles or, equivalently, the invariance of the Hamiltonian under
the $U(1)$ rotation (\ref{16}), (\ref{16p}) (according to the
Noether theorem, such an invariance leads to the conservation of the
number of particles).  We now verify this idea for an ideal gas
using an approach similar to Bogoliubov's, but allowing for a more
general structure of the operator $\hat{\acute{a}}_{0}$.

For the unperturbed Hamiltonian%
\begin{equation}
\hat{H}=\sum_{\mathbf{k}\neq0}\frac{\hbar^{2}k^{2}}{2m}\hat{a}_{\mathbf{k}%
}^{+}\hat{a}_{\mathbf{k}},\label{4-h}%
\end{equation}
the ground state of an infinite system of spinless bosons can be described by
one of the following formulae:%
\begin{equation}
|0\rangle=(N!)^{-1/2}[\hat{a}_{0}^{+}]^{N\rightarrow\infty}|0_{\mathrm{bare}%
}\rangle,\label{4-9}%
\end{equation}%
\begin{equation}
|0\rangle=\sum\limits_{j=0}^{\infty}c_{j}(j!)^{-1/2}[\hat{a}_{0}^{+}%
]^{j}|0_{\mathrm{bare}}\rangle,\label{4-10}%
\end{equation}%
\begin{equation}
|0_{\varphi}\rangle=e^{-A^{2}/2}\cdot e^{[|A|e^{i\varphi}\cdot\hat{a}_{0}%
^{+}]}|0_{\mathrm{bare}}\rangle.\label{4-11}%
\end{equation}
Each of them describes the ground state of the system:
$\hat{H}|0\rangle=0$. The possibility of using several different
formulae is due to the indefiniteness of the number of particles for
an infinite system.

Bogoliubov considered an ideal gas with the Hamiltonian
$\hat{H}^{\prime}$, consisting of the unperturbed part (\ref{4-h})
and a small additional term \cite{bogquasi1,bogquasi2}
\begin{equation}
\delta \hat{H}=-\lambda\sum_{\mathbf{k}}
\hat{a}^{+}_{\mathbf{k}}\hat{a}_{\mathbf{k}}
-\nu\sqrt{V}(\hat{a}^{+}_{0}e^{i\varphi}+\hat{a}_{0}e^{-i\varphi}),
 \label{4-dhb} \end{equation}
where $\nu>0$ is a small parameter ($\nu\rightarrow 0$),
$\lambda=-\nu/ \sqrt{n_{0}}$ (if $T=0$, then $N_{0}=N$ and
$n_{0}=n$), and $\varphi$ is a fixed angle. Such a Hamiltonian can
be written in diagonal form \cite{bogquasi1,bogquasi2}
\begin{eqnarray}
 \hat{H}^{\prime}&=&
\frac{\nu}{\sqrt{n_{0}}}\hat{\acute{a}}^{+}_{0}\hat{\acute{a}}_{0}+
\sum_{\mathbf{k}\neq 0}\left
(\frac{\hbar^{2}k^{2}}{2m}+\frac{\nu}{\sqrt{n_{0}}}\right )
\hat{a}^{+}_{\mathbf{k}}\hat{a}_{\mathbf{k}}\nonumber
\\&-&\nu\sqrt{n_{0}}V,
 \label{4-hb} \end{eqnarray}
with
\begin{equation}
\hat{\acute{a}}_{0}= \hat{a}_{0}-\sqrt{N_{0}}e^{i\varphi}.
 \label{4-a0b} \end{equation}
The
ground-state wave function can be found from the equations $\hat{a}_{\mathbf{k}%
\neq0}|0_{\varphi}\rangle=0$,
$\hat{\acute{a}}_{0}|0_{\varphi}\rangle
\equiv(\hat{a}_{0}-N_{0}^{1/2}e^{i\varphi})|0_{\varphi}\rangle=0$.
Their solution is given by the coherent state (\ref{4-11}) with
$A=N_{0}^{1/2}$ \cite{petrina}. The energy $E_{0}$ of the ground
state can be obtained from the equation
$\hat{H}^{\prime}|0_{\varphi}\rangle=E_{0}|0_{\varphi}\rangle$ and
is equal to $E_{0}=-\nu\sqrt{n_{0}}V$, so that
$E_{0}/N=-\nu\sqrt{n_{0}}/n=-\nu/\sqrt{n}$. Taking the limits
$\nu\rightarrow0$ and $N,V\rightarrow\infty$
\cite{bogquasi1,bogquasi2}, we obtain an ill-defined total energy
$E_{0}\rightarrow\mathrm{const}\cdot0\cdot\infty$, and a
well-defined energy per particle, $E_{0}/N\rightarrow 0$.

The additional term $\delta \hat{H}(\varphi)$ (\ref{4-dhb})
transforms the ground state (\ref{4-9}) into the coherent state
$|0_{\varphi}\rangle$~(\ref{4-11}); in this case,  $\langle 0
|\hat{\psi}(\mathbf{r},t)|0\rangle_{q}=\langle 0_{\varphi}
|\hat{\psi}(\mathbf{r},t)|0_{\varphi}\rangle
=\sqrt{n_{0}}e^{i\varphi}\equiv \sqrt{n}e^{i\varphi}$, which
corresponds to the condensation of all atoms in the zero-momentum
state. The phase $\varphi$ in Eq.~(\ref{4-11}) can be arbitrary;
therefore, this state is infinitely degenerate. Since
$\hat{U}_{\theta}|0_{\varphi}\rangle= |0_{\varphi+\theta}\rangle$,
the ground state (\ref{4-11}) is not invariant under the $U(1)$
rotation. Because the unperturbed Hamiltonian (\ref{4-h}) is
invariant under the $U(1)$ rotation, and the correction $\delta
\hat{H}(\varphi)$ (\ref{4-dhb}) is arbitrarily small and does not
change the energy per atom ($E_{0}/N=-\nu/\sqrt{n}\rightarrow 0$ for
any $\varphi$), we have infinite degeneracy of the ground state and
spontaneous breaking of the $U(1)$ symmetry. In this case, the
introduction of $\delta\hat{H} (\varphi)$ destroys the $U(1)$
invariance of the Hamiltonian and lifts the degeneracy of the ground
state. Therefore, at first glance, it seems natural to conclude that
the degeneracy is related to the $U(1)$ symmetry of the Hamiltonian.
Bogoliubov likewise related the degeneracy to this symmetry
\cite{bogquasi1,bogquasi2}.

However, a more detailed study leads to a different upshot. This is
already evident  from the group-theoretical analysis above. To see
this more clearly, let us carry out a specific analysis. Instead of
Bogoliubov's Hamiltonian (\ref{4-hb}) with $\hat{\acute{a}}_{0}$
(\ref{4-a0b}), let us consider the Hamiltonian
\begin{eqnarray}
&& \hat{H}^{\prime}=
\frac{\nu}{\sqrt{n_{0}}}\hat{\acute{a}}^{+}_{0}\hat{\acute{a}}_{0}+
\sum_{\mathbf{k}\neq 0}\frac{\hbar^{2}k^{2}}{2m}
\hat{a}^{+}_{\mathbf{k}}\hat{a}_{\mathbf{k}}
 \label{4-hme} \end{eqnarray}
with $\nu\rightarrow 0$ and $\hat{\acute{a}}_{0}$ of a more general
form,
\begin{equation}
\hat{\acute{a}}_{0}= b_{0}+b_{1}\hat{a}_{0}+d_{1}\hat{a}^{+}_{0},
 \label{4-a0p1} \end{equation}
where $b_{0}=|b_{0}|e^{i\varphi_{0}}$,
$b_{1}=|b_{1}|e^{i\varphi_{1}}$, and
$d_{1}=|d_{1}|e^{i\varphi_{2}}$.  We thus pass from the operators
$\hat{a}_{0}$ and $\hat{a}^{+}_{0}$ to $\hat{\acute{a}}_{0}$ and
$\hat{\acute{a}}^{+}_{0}$.  We have also replaced the quantity
$-\nu\sqrt{n_{0}}V$ in the Hamiltonian (\ref{4-hb}) with zero, since
this removes the ambiguity in the total energy $E_{0}$:  now
$E_{0}=0$.

Furthermore, the operators $\hat{\acute{a}}_{0}$ and
$\hat{\acute{a}}^{+}_{0}$ must satisfy the bosonic commutation
relations. The relations
$\hat{a}_{\mathbf{k}}\hat{\acute{a}}_{0}-\hat{\acute{a}}_{0}\hat{a}_{\mathbf{k}}=0$
and
$\hat{a}_{\mathbf{k}}\hat{\acute{a}}^{+}_{0}-\hat{\acute{a}}^{+}_{0}\hat{a}_{\mathbf{k}}=0$
are satisfied automatically, while the relation
$\hat{\acute{a}}_{0}\hat{\acute{a}}^{+}_{0}-\hat{\acute{a}}^{+}_{0}\hat{\acute{a}}_{0}=1$
yields the formula
\begin{equation}
|b_{1}|^{2}=|d_{1}|^{2}+1.
 \label{4-16} \end{equation}
Thus, the quantities  $|b_{0}|$, $|d_{1}|$, $\varphi_{0}$,
$\varphi_{1}$, and $\varphi_{2}$ in Eq.  (\ref{4-a0p1}) are
arbitrary numbers, and the value of $|b_{1}|$ is given by formula
(\ref{4-16}).

The ground state $|0\rangle$ is specified by the equations
$\hat{a}_{\mathbf{k}\neq 0}|0\rangle=0$ and
$\hat{\acute{a}}_{0}|0\rangle  = 0$. Such a state can only be
constructed using the operators $\hat{a}^{+}_{0}$. The general
solution of this type has the form (\ref{4-10}). In this case, the
equation $\hat{a}_{\mathbf{k}\neq 0}|0\rangle=0$ is automatically
satisfied, while $\hat{\acute{a}}_{0}|0\rangle  = 0$ yields the
equation
\begin{eqnarray}
&&
(b_{0}+b_{1}\hat{a}_{0}+d_{1}\hat{a}^{+}_{0})\sum\limits_{j=0}^{\infty}
\frac{c_{j}}{\sqrt{j!}}[\hat{a}^{+}_{0}]^{j}|0_{bare}\rangle=0.
 \label{4-17} \end{eqnarray}
Using the formula
\begin{eqnarray}
&&
\hat{a}_{0}(\hat{a}^{+}_{0})^{j}|0_{bare}\rangle=j(\hat{a}^{+}_{0})^{j-1}|0_{bare}\rangle,
\quad j\geq 1,
 \label{4-18} \end{eqnarray}
one can show that Eq. (\ref{4-17}) is satisfied when
\begin{equation}
c_{1}=-\frac{b_{0}c_{0}}{b_{1}},
 \label{4-19} \end{equation}
\begin{equation}
c_{j+1}=-\frac{b_{0}c_{j}+d_{1}\sqrt{j}\cdot
c_{j-1}}{b_{1}\sqrt{j+1}}, \quad j\geq 1.
 \label{4-20} \end{equation}
Formulae (\ref{4-19}) and (\ref{4-20}) allow one to express any
coefficient $c_{j\geq 1}$ in terms of $c_{0}$. In this way, we have
found the state $|0\rangle$ (\ref{4-10}) up to an arbitrary constant
factor $c_{0}$.

Let us find the quasi-average $C=\langle
0|\hat{\psi}(\mathbf{r},t)|0\rangle_{q}$. Using relations (\ref{3}),
(\ref{4-18}) and
\begin{equation}
\langle N_{0}|N_{0}\rangle =1, \quad |N_{0}\rangle
=(\hat{a}^{+}_{0})^{N_{0}}(N_{0}!)^{-1/2}|0_{bare}\rangle,
 \label{4-21} \end{equation}
we obtain
\begin{equation}
\langle 0| \hat{\psi}(\mathbf{r})|0\rangle_{q}
=\frac{1}{\sqrt{V}}\sum_{j=0}^{\infty} \sqrt{j+1}c^{*}_{j}c_{j+1}.
 \label{4-23} \end{equation}
By means of Eqs. (\ref{4-19}) and (\ref{4-20}), formula (\ref{4-23})
is reduced to
\begin{equation}
C=-\frac{b_{0}}{b_{1}\sqrt{V}}\sum\limits_{j=0}^{\infty}
|c_{j}|^{2}- \frac{d_{1}}{b_{1}}C^{*}.
 \label{4-24a} \end{equation}
Using the relation
\begin{equation}
\sum\limits_{j=0}^{\infty} |c_{j}|^{2}=1,
 \label{4-22} \end{equation}
which follows from the normalisation $\langle 0|0\rangle =1$ and
relations (\ref{4-21}),  we can write Eq. (\ref{4-24a}) as
\begin{equation}
C=-\frac{b_{0}}{b_{1}\sqrt{V}} - \frac{d_{1}}{b_{1}}C^{*},
 \label{4-24} \end{equation}
\begin{equation}
C^{*}=-\frac{b^{*}_{0}}{b^{*}_{1}\sqrt{V}} -
\frac{d^{*}_{1}}{b^{*}_{1}}C. \label{4-25} \end{equation}%
From (\ref{4-16}), (\ref{4-24}) and (\ref{4-25}) we finally find
that
\begin{eqnarray}
&&C\equiv\langle 0| \hat{\psi}(\mathbf{r})|0\rangle_{q}
=\frac{-b_{0}b^{*}_{1}+d_{1}b^{*}_{0}}{\sqrt{V}}\nonumber \\&&=
\frac{|b_{0}|}{\sqrt{V}}\left
[-|b_{1}|e^{i(\varphi_{0}-\varphi_{1})}+|d_{1}|e^{i(\varphi_{2}-\varphi_{0})}\right
].
\label{4-26} \end{eqnarray}%
The value of $\langle 0| \hat{\psi}(\mathbf{r})|0\rangle_{q}$ is
nonzero as a consequence of particle-number indeterminacy in the
ground state (\ref{4-10}). For $b_{0}=-\sqrt{N_{0}}e^{i\varphi}$,
$b_{1}=1$, and $d_{1}=0$, our formulae  (\ref{4-a0p1}) and
(\ref{4-26}) become Bogoliubov's formulae (\ref{4-a0b}) and $\langle
0| \hat{\psi}(\mathbf{r})|0\rangle_{q}=\sqrt{n_{0}}e^{i\varphi}$.
Note that the quasi-average (\ref{4-26}) is independent of the small
parameter $\nu $ (recall that we are considering the case of
$T\rightarrow 0$).

 Let us revert to Bogoliubov's idea according to which statistical
degeneracy (and therefore ground-state degeneracy) is related to the
$U(1)$ symmetry of the Hamiltonian. Our additional term $\delta
\hat{H}=\frac{\nu}{\sqrt{n_{0}}}\hat{\acute{a}}^{+}_{0}\hat{\acute{a}}_{0}$
explicitly breaks the $U(1)$ symmetry of the Hamiltonian
(\ref{4-hme}) and leads to infinite degeneracy of the ground state
(\ref{4-10}), inasmuch as the values of $|b_{0}|$, $|d_{1}|$,
$\varphi_{0}$, $\varphi_{1}$, and $\varphi_{2}$ are arbitrary.
Moreover, instead of (\ref{4-a0p1}), one can consider an expansion
of general form,
\begin{eqnarray}
\hat{\acute{a}}_{0}&=&
\sum\limits_{p,j=0}^{\infty}b_{pj}(\hat{a}^{+}_{0})^{p}\hat{a}^{j}_{0}=b_{00}+b_{01}\hat{a}_{0}+b_{10}\hat{a}^{+}_{0}\nonumber
\\ &+&
b_{02}\hat{a}^{2}_{0}+b_{11}\hat{a}^{+}_{0}\hat{a}_{0}+b_{20}(\hat{a}^{+}_{0})^{2}+\ldots,
 \label{4-27} \end{eqnarray}
where $b_{pj}=|b_{pj}|e^{i\varphi_{pj}}$. Normalisation $\langle
0|0\rangle =1$ imposes a limitation (\ref{4-22}), and the
commutation relation
$\hat{\acute{a}}_{0}\hat{\acute{a}}^{+}_{0}-\hat{\acute{a}}^{+}_{0}\hat{\acute{a}}_{0}=1$
must result in connections between the various coefficients
$b_{pj}$. Despite these limitations, we may expect that an infinite
number of $|b_{pj}|$ and $\varphi_{pj}$ in  formula (\ref{4-27})
will still be arbitrary. Each of these parameters gives rise to
infinite degeneracy of the ground state (\ref{4-10}). It is clear
that this degeneracy is not related to the $U(1)$ symmetry of the
Hamiltonian. This shows that there is an additional infinite
degeneracy besides the phase degeneracy in $\varphi$, disclosed by
Bogoliubov.

The source of the infinite degeneracy of the ground state can be
readily identified. The ground state of a \emph{finite} system is
described by the WF (\ref{4-9}) and is non-degenerate. At the same
time, the initial Hamiltonian (\ref{4-h}) is invariant under $U(1)$
transformations. Thus, for a finite system, the $U(1)$ symmetry of
the Hamiltonian does not lead to ground-state degeneracy. However,
as seen from the analysis above, the ground state of an infinite
system with the same Hamiltonian turns out to be degenerate.
Clearly, this degeneracy arises precisely from the transition to an
infinite system.

Moreover, it is well known that
$\langle0|\hat{\psi}(\mathbf{r})|0\rangle\neq0$ also for a
Hamiltonian without $\delta\hat{H}$, provided that $|0\rangle$ is a
state with an indeterminate number of particles. As discussed in
section 2.1, the property $\langle 0|
\hat{\psi}(\mathbf{r})|0\rangle \neq 0$ always leads to ground-state
degeneracy and SSB.

Bogoliubov concluded that the degeneracy is caused by the $U(1)$
symmetry of the Hamiltonian,  because he did not see
that the transition to the thermodynamic limit itself already gives
rise to degeneracy. Furthermore, he appears to have assumed that the
connection between statistical degeneracy and Hamiltonian symmetries
is of a universal character and therefore applies to any continuous
symmetry. Accordingly, he chose the transformation in the form
(\ref{4-a0b}), which corresponds precisely to the $U(1)$ symmetry.
The ground state (\ref{4-11}) that follows from this transformation
is characterised by phase degeneracy in $\varphi$ at fixed $A$. The
connection between degeneracy and the $U(1)$ symmetry is thereby
introduced into the model through this particular choice of
transformation.

However, the ground state is also described by a more general WF
(\ref{4-10}), which contains an infinite number of phases
$\alpha_{j}$ (because $c_{j}=|c_{j}|e^{i\alpha_{j}}$, where
$j=0,1,2,\ldots,\infty$). The WF (\ref{4-10}) can be written as an
expansion in the coherent states,%
\begin{equation}
|0\rangle=\int\limits_{0}^{\infty}A\cdot
dA\int\limits_{0}^{2\pi}d\varphi\cdot
c_{A,\varphi}e^{-A^{2}/2}e^{[|A|e^{i\varphi}\cdot\hat{a}_{0}^{+}%
]}|0_{\mathrm{bare}}\rangle,\label{4-15}%
\end{equation}
since the latter form an overcomplete set of non-orthogonal basis
functions \cite{svidzynsky2004}. In formula (\ref{4-15}),  both $A$
and $\varphi$  (not only $\varphi$) take various values, which
removes the direct association with the $U(1)$ symmetry. In the
method of quasi-averages, the values of $\varphi$ and $A$ in
Eq.~(\ref{4-11}) are set by the choice of $\delta\hat{H}$ (in this
case, the value of $\varphi$ can be arbitrary, whereas $A$ is chosen
to be that obtained in models without quasi-averages).  In general,
however, the values of $\varphi$ and $A$ can be arbitrary.

In this way, our analysis shows that the genuine source of the
ground state degeneracy for Hamiltonians (\ref{4-h}), (\ref{4-hb})
 and (\ref{4-hme}) is the same: it is particle-number
indeterminacy. Indeed, if the number of particles in the state
$|0\rangle$ is certain and equal to $N$, then
$\hat{N}|0\rangle=N|0\rangle$, which implies
$\hat{U}_{\theta}|0\rangle\equiv e^{i\theta\hat{N}}|0\rangle
=e^{i\theta N}|0\rangle$. Such a state $|0\rangle$ is nondegenerate
with respect to $U(1)$.  Therefore, the relation $\hat
{U}_{\theta}|0_{\varphi}\rangle=|0_{\theta+\varphi}\rangle$, which
we have obtained above and which indicates degeneracy, is only
possible when the particle number is indeterminate. The ground state
of an infinite system of spinless \emph{interacting} bosons is
described by the WF (\ref{13}). In a similar way, one can see that
the degeneracy of this state is also related to particle-number
indeterminacy.

Thus, we have shown in several ways that the ground state of an
infinite Bose gas can be considered infinitely degenerate, and that
the degeneracy is related to the indeterminacy of the number of
particles in the infinite system. Although for some systems (e.g., a
ferromagnet), the statistical degeneracy is related to the additive
conservation law \cite{bogquasi1,bogquasi2}, the nature of the
degeneracy for the Bose gas is different. We suppose that the $U(1)$
symmetry of the Hamiltonian does not lead to degeneracy because the
$U(1)$ rotation (\ref{15}) is similar to the wavefunction
transformation $\Psi(\mathbf{r},t)\rightarrow
e^{i\varphi}\Psi(\mathbf{r},t)$ and does not alter the state of the
system.

The existence of degeneracy for the ground state is partly
\textquotedblleft regulated\textquotedblright\ by the
Courant--Hilbert theorem~\cite{gilbert}. This theorem has been
proven for a one-dimensional system of two interacting particles. It
can easily be generalised to the case of a system of any dimension
(1, 2, or 3), with any number of particles ($N\geq2$). According to
the theorem, the ground state of such a system is non-degenerate.
Degeneracy and SSB are possible if the conditions of the theorem are
violated. The violating factors are, in particular, the spin, the
intrinsic multipole moment, the external field, and the infinity of
the system (see \cite{sgs} for more details). For a finite system of
spinless bosons in the absence of an external field, the conditions
of the theorem are satisfied. Therefore, this theorem alone proves
that SSB is impossible for such a system.

The method of quasi-averages makes it possible to study the
stability of the solution with respect to a small perturbation,
$\delta\hat{H}$, which reduces the symmetry of the Hamiltonian.
Therefore, in principle, it can be used to describe equilibrium
states with a symmetry lower than that of the
Hamiltonian~\cite{bogquasi1,bogquasi2,petrina,akhiezer1981}. The
method also allows one to \textquotedblleft
rescue\textquotedblright\ the $c$-number. However, it is worth
noting that operator approaches which do not use the $c$-number are
generally more accurate, both
qualitatively~\cite{gardiner1997,girardeau1998,zagrebnov2007,wu1961,gp1}
and quantitatively~\cite{gp1,gp2}, than approaches that do.

Note also that the method of quasi-averages relies on statistical
degeneracy and, consequently, on the degeneracy of the ground state
(provided that this state belongs to a phase in which the
quasi-average differs from the corresponding average). This implies
that the method is strictly applicable only to systems whose ground
state is degenerate. At the same time, the method is only applicable
to infinite systems and does not, by itself, allow one to ascertain
whether a corresponding finite system is degenerate. Yet any real
physical system is finite. In view of this, the method of
quasi-averages should be applied with caution. In our view, the
problem is addressed in a physically clearer and more reliable
manner if a finite system is considered and the artificially
introduced term $\delta\hat{H}$ is not used (see also section 2.2 in
the monograph \cite{leggett2006}).

\section{Concluding remarks}

In quantum field theory, the ground state is formally structureless
because it is a state \emph{without particles}. Therefore, its
properties can be studied only indirectly: the non-invariance of the
ground state and  SSB are indicated by the non-zero average
$\langle0|\hat{\varphi}|0\rangle\neq0$. In quantum mechanics, the
ground state $|0\rangle$ of a system of $N$ interacting particles is
a state \emph{without quasiparticles}. For this state, not only the
average $\langle0|\hat{\Psi}(\mathbf{r},t)|0\rangle$ can be
calculated but also the quantity $\hat{U}_{\varphi}|0\rangle$, i.e.,
it is possible to directly analyse the $U(1)$-symmetry properties of
the ground state. The main result of this paper is that we have
calculated $\hat{U}_{\varphi }|0\rangle$ for a finite periodic
system of interacting spinless bosons and showed that
$\hat{U}_{\varphi}|0\rangle=e^{iN\varphi}|0\rangle$ and
$\langle0|\hat{\Psi}(\mathbf{r},t)|0\rangle=0$. This means that SSB
is absent in a finite Bose gas or liquid. Therefore, phonons in the
superfluid phase of such a system do not resemble Goldstone bosons.

In some papers, it has been claimed that there is spontaneous
breaking of the $U(1)$ symmetry in a weakly interacting Bose gas and
that, as a consequence, phonons in such a gas are Goldstone bosons.
For a finite system, such a statement is simply a mistake, resulting
from an overly approximate treatment of the problem. Above, we
started from a rigorous definition of SSB and studied the invariance
of $|0\rangle$ under $U(1)$ rotation directly. Using two methods,
one of which is exact, we have demonstrated that SSB is absent in a
finite system.
We have also shown that in the case of an infinite Bose gas, one may
view the system as exhibiting SSB and infinite ground-state
degeneracy; at the same time, one may regard the degeneracy and
spontaneous symmetry breaking as absent. This duality is related to
the paradoxical properties of infinity.
In this case, the infinite degeneracy is rooted in the indeterminacy
of the particle number $N$ at~$N=\infty$, rather than in the $U(1)$
invariance of the Hamiltonian, as is commonly believed.

Note that our conclusions regarding finite systems are valid for any
space dimension (1, 2, and 3), since the formulae in Section 2.3 are
applicable in the case of any dimension. In some books, one can read
that the Bose condensation of atoms, crystalline ordering, and SSB
are impossible in one dimensions (1D) and two dimensions (2D). This
is true only for infinite systems; for finite systems, those
properties are possible in 1D and 2D, as shown in a number of works.
These features, together with the results of this article, imply
that the transition to the thermodynamic limit---widely used in
physics---can lead to physically misleading conclusions when applied
to real systems, \emph{which are always finite}. Although in most
cases such a transition is completely justified. In particular, the
value $N_{0}/N$ of atomic condensate (in 3D), the ground state
energy $E_{0}/N$, and the quasiparticle dispersion law are the same
for a Bose gas of very large finite volume and of infinite volume
(at the same particle density; see the results for finite
\cite{holes2020,gp1,mtsp2019,cazalilla2004,mt2015,mtmethodbog} and
infinite \cite{yuv2,yuv1,ll1963,lieb1963,vak1989,vak1990,mt2006}
systems, as well as the results
\cite{bog1947,bz,gardiner1997,girardeau1998,zagrebnov2007,girardeau1960},
that are valid for both finite and infinite systems).

Another important point. Bogoliubov's $1/q^2$-theorem
\cite{bogquasi1,bogquasi2} is considered to be similar to the
Goldstone theorem \cite{miransky1994}. According to the
$1/q^2$-theorem, the gapless nature of the dispersion law of a Bose
gas is related to the spontaneous breakdown of the $U(1)$ symmetry.
This theorem is only valid for an infinite system. However, we noted
in the previous item that the dispersion law of a Bose gas does not
change at the transition to the thermodynamic limit (this follows,
in particular, from \emph{exact} solutions
\cite{holes2020,mtsp2019,lieb1963}). That is, the dispersion law of
a finite system  is also gapless; however, SSB is absent in this
case.
This suggests that the gapless character of the dispersion law of
the infinite system is not a consequence of SSB, although the system
does exhibit SSB. If this is the case, then the similarity between
the Goldstone theorem and the $1/q^2$-theorem is only formal. This
question requires an additional study.

Our results also shed light on the nature of superfluidity in a
system of spinless bosons. According to the widely accepted view,
superfluidity is related to a condensate of atoms and the fulfilment
of Landau's criterion for quasiparticles. Some authors believe that
the condensate implies a spontaneous breaking of the $U(1)$
symmetry, while others remain silent on the matter. If phonons were
similar to Goldstone bosons, then the fulfilment of Landau's
criterion would also be related to the spontaneous breaking of the
$U(1)$ symmetry. Thus, if SSB were present, it would be the original
cause of superfluidity. However, we have shown above that the
spontaneous breakdown of the $U(1)$ symmetry is absent in a
\emph{finite} system of spinless bosons.  Consequently, \emph{in
real-world systems, the condensate of atoms and superfluidity are
entirely unrelated
to the breaking of the $U(1)$ symmetry}, and phonons have the same
nature at temperatures below and above $T_{\lambda}$: they exist due
to the interaction between atoms and are not related  to Goldstone
bosons. The latter property is also evidenced by the closeness of
the profile of the $^{4}$He structure factor $S(k,\omega)$ for
$T=T_{\lambda}-\delta$ to the profile for $T=T_{\lambda}+\delta$,
where
$0<\delta\ll T_{\lambda}$%
~\cite{andersen1994a,andersen1994b,blag1997,andersen1999,kalinin2007}.
The quasiparticle dispersion laws for liquid $^{4}$He at $T=T_{\lambda}%
-\delta$ and $T=T_{\lambda}+\delta$ are close and satisfy the Landau
criterion.

This study was inspired by the monograph of V. Miransky
\cite{miransky1994}, in which we happened to find
a mathematically accurate approach for studying the problem of
spontaneous $U(1)$ symmetry breaking in quantum-mechanical
many-particle systems.

\section*{Acknowledgements}

The author is grateful for the financial support from the National
Academy of Sciences of Ukraine (project No.~0126U000353) and the
Simons Foundation International (grant SFI-PD-Ukraine-00014573, PI
LB).


\appendix

\section*{Appendix}
\setcounter{section}{1}

Consider a relation between the degeneracy of energy levels and the
symmetry of the Hamiltonian \cite{elliott,petrashen} because this is
the key point of our analysis.

Let the boundary value problem (the Hamiltonian $\hat{H}$ and the
BCs) be invariant under a group $G$. Let $g$ be an element of $G$,
and let the operators $\hat{T}(g)$ form a group isomorphic to $G$
(if the symmetry of the Hamiltonian is lower than that of the BCs,
or vice versa, then $G$ should be chosen as a group with respect to
which $\hat{H}$
and the BCs are invariant). Then%
\begin{equation}
\hat{T}(g_{1})\hat{T}(g_{2})=\hat{T}(g_{1}g_{2}), \label{a-1}%
\end{equation}%
\begin{equation}
\lbrack\hat{H},\hat{T}(g)]=0. \label{a-2}%
\end{equation}
The function $\hat{T}(g)\Psi_{j}$ can be expanded in the complete
set of
eigenfunctions, $\{\Psi_{l}\}$, of the Hamiltonian~$\hat{H}$: $\hat{T}%
(g)\Psi_{j}=\sum_{l}T_{lj}(g)\Psi_{l}$. It is easy to show that the matrices
$T_{lj}(g)$ define the group representation,%
\begin{equation}
T(g_{1})T(g_{2})=T(g_{1}g_{2}). \label{a-3}%
\end{equation}
Indeed, using formula (\ref{a-1}), we obtain%
\begin{eqnarray}
&&\hat{T}(g_{1})\hat{T}(g_{2})\Psi_{j}=\hat{T}(g_{1})\sum_{l}T_{lj}(g_{2}%
)\Psi_{l}\nonumber
\\&&=\sum_{lp}T_{lj}(g_{2})T_{pl}(g_{1})\Psi_{p}\nonumber
\\&&=\sum_{p}\left( \sum
_{l}T_{pl}(g_{1})T_{lj}(g_{2})\right)  \Psi_{p}. \label{a-4}%
\end{eqnarray}
On the other hand,%
\begin{equation}
\hat{T}(g_{1})\hat{T}(g_{2})\Psi_{j}=\hat{T}(g_{1}g_{2})\Psi_{j}=\sum
_{p}T_{pj}(g_{1}g_{2})\Psi_{p}. \label{a-5}%
\end{equation}
From Eqs.~(\ref{a-4}) and (\ref{a-5}), it follows that $T_{pj}(g_{1}%
g_{2})=\sum_{l}T_{pl}(g_{1})T_{lj}(g_{2})$, i.e. formula
(\ref{a-3}).

If the representation $T_{lj}(g)$ is unitary, then with the help of a linear
transformation, the basis functions $\Psi_{j}$ can be reduced to a form where
the representation $T_{lj}(g)$ is a set of irreducible
representations~\cite{elliott,petrashen}. Then%
\begin{equation}
\hat{T}(g)=\hat{T}^{(1)}(g)\oplus\hat{T}^{(2)}(g)\oplus\ldots\oplus\hat
{T}^{(\kappa)}(g), \label{a-6}%
\end{equation}
where $\kappa$ is the number of irreducible representations. In this case, if
the functions $\Psi_{j}^{(l)}$ are the basis functions of the $l$-th
irreducible representation, then $\hat{T}(g)\Psi_{j}^{(l)}=\hat{T}%
^{(l)}(g)\Psi_{j}^{(l)}=\sum_{p=1}^{P_{l}}T_{pj}^{(l)}(g)\Psi_{p}^{(l)}$
for any $g$; here $P_{l}$ is the dimension of the $l$-th irreducible
representation.

Let the eigenfunctions $\Psi_{j}^{[p]}$ of the Hamiltonian
correspond to the eigenenergies $E_{p}$:
$E_{0},E_{1},\ldots,E_{\infty}$. In this case let the eigenfunctions
$\Psi_{j=1,\ldots,J_{l}}^{[l]}$ correspond to the same energy
$E_{l}$, i.e.,
\begin{equation}
\hat{H}\Psi_{j}^{[l]}=E_{l}\Psi_{j}^{[l]}. \label{1-new00}%
\end{equation}
Making use of Eqs. (\ref{a-2}) and (\ref{1-new00}), we get
\begin{eqnarray}
E_{l}\hat{T}(g)\Psi_{j}^{[l]}&=&\hat{T}(g)E_{l}\Psi_{j}^{[l]}=\hat{T}(g)\hat
{H}\Psi_{j}^{[l]}\nonumber \\&=&\hat{H}\hat{T}(g)\Psi_{j}^{[l]}. \label{1-new000}%
\end{eqnarray}
So, for any $j=1,\ldots,J_{l}$ the function
$\hat{T}(g)\Psi_{j}^{[l]}$ is also an eigenfunction of the
Hamiltonian with the energy $E_{l}$. This means that for each
$j=1,\ldots,J_{l}$ the function $\hat{T}(g)\Psi_{j}^{[l]}$ can be
written in the form
$\sum_{p=1}^{J_{l}}c_{pj}^{[l]}(g)\Psi_{p}^{[l]}$. Therefore, the
functions $\Psi_{j=1,\ldots,J_{l}}^{[l]}$  transform according to
the representation of the group $G$. This representation is
irreducible if each irreducible representation of this group
corresponds to its own specific energy value. In this case, the
functions $\Psi _{j=1,\ldots,J_{l}}^{[l]}$ can be chosen as the
basis functions of the $l$-th irreducible representation:
$\Psi_{j}^{[l]}=\Psi_{j}^{(l)}$, $J_{l}=P_{l}$. If the energy
$E_{l}$ corresponds to several irreducible representations (this
happens rarely and is called accidental degeneracy), then the
functions $\Psi_{j=1,\ldots,J_{l}}^{[l]}$ transform according to the
representation of the group $G$, which is reduced to these
irreducible representations. These properties mean that (i)~the
eigenfunctions of the Hamiltonian can be chosen in such a way that
they transform according to the irreducible representations of the
symmetry group $G$ of the Hamiltonian, and (ii)~if there is no
accidental degeneracy, then the degeneracy multiplicity of the state
with the energy $E_{l}$ is equal to the dimensionality of the $l$-th
irreducible representation. Such an analysis is applicable to a
continuous symmetry of any type, i.e., both intrinsic and spatial.

\end{document}